\documentclass{aa}  
\usepackage{graphicx}
\usepackage{txfonts}
\usepackage{natbib}
\bibpunct{(}{)}{;}{a}{}{,} 

\newcommand{\ec}{EC~23487-2424}
\newcommand{\bpm}{BPM~31594}
\newcommand{\hs}{HS~0507+0434B}
\newcommand{\ross}{Ross~548}
\newcommand{\bpmb}{BPM~30551}
\newcommand{\mct}{MCT~0145-2211}
\newcommand{\he}{HE~0532-5605}
\newcommand{\lnt}{L~19-2}
\usepackage{xcolor}

\usepackage{soul} 
\definecolor{rr}{HTML}{e67e22}

\begin{document}

   \title{\textit{TESS} first look at evolved compact pulsators:}

   \subtitle{Known ZZ~Ceti stars of the southern ecliptic hemisphere as seen by \textit{TESS}}

    
    \author{Zs.~Bogn\'ar\inst{1,2,3}\fnmsep\thanks{\email{bognar@konkoly.hu}},
        S.~D.~Kawaler\inst{4},
        K.~J.~Bell\inst{5,6},
        C.~Schrandt\inst{4},
        A.~S.~Baran\inst{7},
        P.~A.~Bradley\inst{8},
        J.~J.~Hermes\inst{9},
        S.~Charpinet\inst{10},
        G.~Handler\inst{11},
        S.~E.~Mullally\inst{12},
        S.~J.~Murphy\inst{13},
        R.~Raddi\inst{14,15},
        \'A.~S\'odor\inst{1,2},
        P.-E.~Tremblay\inst{16}
        M.~Uzundag\inst{17},
        \and W.~Zong\inst{18}}
   
   \institute{
        Konkoly Observatory, Research Centre for Astronomy and Earth Sciences, Konkoly Thege Mikl\'os \'ut 15-17, H--1121, Budapest, Hungary
        \and
        MTA CSFK Lend\"ulet Near-Field Cosmology Research Group
        \and
        ELTE E\"otv\"os Lor\'and University, Institute of Phyiscs, P\'azm\'any P\'eter s\'et\'any 1/A, H-1171, Budapest, Hungary
        \and
        Department of Physics and Astronomy, Iowa State University, Ames, IA 50011, USA
        \and
        DIRAC Institute, Department of Astronomy, University of Washington, Seattle, WA 98195-1580, USA
        \and
        NSF Astronomy and Astrophysics Fellow and DIRAC Fellow
        \and
        ARDASTELLA Research Group, Institute of Physics, Pedagogical University of Cracow, ul. Podchor\c{a}\.zych 2, 30-084 Krak\'ow, Poland
        \and
        XCP-6, MS F-699 Los Alamos National Laboratory, Los Alamos, NM 87545, USA
        \and
        Department of Astronomy, Boston University, 725 Commonwealth Ave., Boston, MA 02215 - USA
        \and
        Institut de Recherche en Astrophysique et Plan\'etologie, CNRS, Universit\'e de Toulouse, CNES, 14 avenue Edouard Belin, F-31400 Toulouse, France
        \and
        Nicolaus Copernicus Astronomical Center, Bartycka 18, 00-716 Warsaw, Poland
        \and
        Space Telescope Science Institute, 3700 San Martin Dr., Baltimore MD 21212
        \and
        Sydney Institute for Astronomy (SIfA), School of Physics, The University of Sydney, NSW 2006, Australia
        \and
        Dr. Remeis-Sternwarte, Friedrich-Alexander Universit\"at Erlangen-N\"urnberg, Sternwartstr. 7, 96049 Bamberg, Germany
        \and
        Universitat Politecnica de Catalunya, Departament de Fisica, c/ Esteve Terrades, 5, 08860 Castelldefels, Spain
        \and
        Department of Physics, University of Warwick, Coventry CV4 7AL, UK
        \and
        Instituto de F\'isica y Astronom\'ia, Universidad de Valparaiso, Gran Breta\~na 1111, Playa Ancha, Valpara\'iso 2360102, Chile
        \and
        Department of Astronomy, Beijing Normal University, Beijing 100875, China
        }
        
    \titlerunning{Known ZZ~Ceti stars of the southern ecliptic hemisphere as seen by \textit{TESS}}
	\authorrunning{Zs.~Bogn\'ar et al.}
        
    \date{}

 
  \abstract
   {We present our findings on 18 formerly known ZZ~Ceti stars observed by the \textit{TESS} space telescope in 120\,s cadence mode during the survey observation of the southern ecliptic hemisphere.}
   {We focus on the frequency analysis of the space-based observations, comparing the results with the findings  of  the  previous  ground-based measurements. The frequencies detected by the \textit{TESS} observations can serve as inputs for future asteroseismic analyses.}
   {We performed standard pre-whitening of the data sets to derive the possible pulsation frequencies of the different targets. In some cases, we fitted Lorentzians to the frequency groups that emerged as the results of short-term amplitude/phase variations that occurred during the \textit{TESS} observations.}
  {We detected more than 40 pulsation frequencies in seven ZZ~Ceti stars observed in the 120\,s cadence by \textit{TESS}, with better than $0.1\,\mu$Hz precision. We found that HE~0532$-$5605 may be a new outbursting ZZ~Ceti. Ten targets do not show any significant pulsation frequencies in their Fourier transforms, due to a combination of their intrinsic faintness and/or crowding on the large \textit{TESS} pixels. We also detected possible amplitude/phase variations during the \textit{TESS} observations in some cases. Such behaviour in these targets was not previously identified from ground-based observations.}
   {}


   \maketitle
%

\section{Introduction}

A new era started in the field of variable star studies on 18 April 2018, when \textit{TESS} (\textit{Transiting Exoplanet Survey Satellite}; \citealt{2015JATIS...1a4003R}) was launched successfully. As part of NASA’s Explorer programme, the main goal of this almost all-sky survey space mission is to detect exoplanets at nearby and bright stars with the transit method. However, with the time sampling of 30\,min of the Full-Frame Images (FFIs) provided by the satellite, and with the 120\,s short-cadence mode available for selected targets, it is also possible to study the light variations of different classes of pulsating variable stars, including the pulsations of the short-period compact variables: (pre-)white dwarf and hot subdwarf stars. These later activities are coordinated by the \textit{TESS} Asteroseismic Science Consortium (TASC) Compact Pulsators Working Group (WG\#8).

The most populous group of pulsating white dwarf stars is also intrinsically the least luminous: the hydrogen-atmosphere pulsating white dwarfs (DAVs), named after their prototype ZZ~Ceti. These stars are short-period (P\,$\sim$\,100--1500\,s), low-amplitude (\textit{A}\,$\sim$\,0.1\%) pulsators with effective temperatures in the range of $T_{\rm eff} =$ 10\,500--13\,000\,K. Pulsation modes detected in these objects are low spherical degree ($\ell = 1$ and $2$), low-to-mid radial order $g$-modes. 

A known characteristic of the ZZ~Ceti pulsations is that we detect different pulsational behaviour within the temperature range that defines their instability strip. While variables closer to the blue edge (higher $T_{\rm eff}$) are more likely to show pulsational frequencies with stable amplitudes and phases, short-term (days--weeks) amplitude and phase changes are more common closer to the red edge (lower $T_{\rm eff}$). Longer periods and larger amplitudes are also detected at the cooler objects. The short-term amplitude variations can be results of the insufficient frequency resolution of the data sets, possible physical explanations are interaction of pulsation and convection (e.g. \citealt{2010ApJ...716...84M}), and resonant mode coupling (e.g. \citealt{2016A&A...585A..22Z}).

The \textit{Kepler} space-telescope \citep{2010ApJ...713L..79K} observations revealed another unusual behaviour of these objects recently, the so-called outburst events: recurring increases in the stellar flux (up to 15 per cent) in cool ZZ~Ceti stars (see e.g. \citealt{2017ASPC..509..303B}). This phenomenon might be in connection with the cessation of pulsations at the empirical red edge of the ZZ~Ceti instability strip \citep{2015ApJ...810L...5H}. 

For comprehensive reviews of the observational and theoretical aspects of pulsating white dwarf studies, see the papers of \citet{2008ARA&A..46..157W}, \citet{2008PASP..120.1043F}, \citet{2010A&ARv..18..471A}, and \citet{2019arXiv190700115C}. For a study on the pulsational properties of ZZ~Ceti stars based on mainly space-based observations, see the paper of \citet{2017ApJS..232...23H}.

This paper is part of a series presenting the first results on compact pulsators based on \textit{TESS} measurements. The other papers of this series focus on different types of compact variables: an sdBV (hot B subdwarf pulsator) star \citep{2019arXiv191004234C}, another three sdBV stars (Sahoo et al., in prep.), a DBV (pulsating helium-atmosphere white dwarf) \citep{2019A&A...632A..42B} and a DOV (hot hydrogen-deficient post-Asymptotic Giant Branch) variable (Sowicka et al., in prep).

This manuscript presents the study of 18 formerly known ZZ~Ceti stars bright enough to be targeted with 2-min cadence observations with TESS, observed during the survey of the southern ecliptic hemisphere. We present the 120\,s cadence-mode \textit{TESS} measurements collected on these stars in Sect.~\ref{sect:tess}, give details on the frequency analysis of the data in Sect.~\ref{sect:analyses} both for the stars with detected light variations and those not observed to vary this time, and we discuss our findings in Sect.~\ref{sect:disc}.   

\section{\textit{TESS} observations}
\label{sect:tess}

\begin{table*}
\centering
\caption{Journal of observations of the eight targets showing light variations in their \textit{TESS} data sets. All data were collected in 120\,s-cadence mode. \textit{N} is the number of data points after the removal of all those with quality warning flags, $\delta T$ is the length of the data sets including gaps, and \textit{Sect.} is the serial number of the sector(s) in which the star was observed. The start time in BJD is the time of the first data point in the reduced data set. The \textit{CROWDSAP} keyword represents the ratio of the target flux to the total flux in the \textit{TESS} aperture.}
\label{tabl:journal1}
\begin{tabular}{lccrrcrcr}
\hline
\hline
Object & TIC & Start time & \multicolumn{1}{c}{\textit{N}} & \multicolumn{1}{c}{$\delta T$} & \textit{TESS} mag & Sect. & \multicolumn{1}{c}{CROWDSAP} & \multicolumn{1}{c}{$0.1\%$\,FAP} \\
& & (BJD-2\,457\,000) & & \multicolumn{1}{c}{(d)} & & & & \multicolumn{1}{c}{(mma)}\\
\hline
\ross\ & 029854433 & 1385.954 & 13\,452 & 20.3 & 14.3 & 3 & 0.97 & 1.24\\
\ec\ & 033986466 & 1354.115 & 18\,316 & 27.4 & 15.4 & 2 & 0.86 & 2.83\\
\bpm\ & 101014997 & 1385.951 & 29\,244 & 50.9 & 15.1 & 3--4 & 0.93 & 1.89\\
\bpmb\ & 102048288 & 1354.113 & 18\,311 & 27.4 & 15.5 & 2 & 0.96 & 3.13\\
\mct\ & 164772507 & 1385.953 & 13\,448 & 20.3 & 15.2 & 3 & 0.98 & 3.00\\
\lnt\ & 262872628 & 1624.959 & 19\,103 & 27.9 & 13.5 & 12 & 0.80 & 0.65\\
\he\ & 382303117 & 1325.295 & 213\,592 & 357.1 & 16.0 & 1--13 & 0.79 & 1.93\\
\hs\ & 455094688 & 1437.997 & 17\,626 & 26.0 & 15.4 & 5 & 0.40 & 4.23\\
\hline
\end{tabular}
\end{table*}

\begin{table*}
\centering
\caption{Journal of observations of the targets not observed to vary by the \textit{TESS} data sets.}
\label{tabl:journal2}
\begin{tabular}{lccrrcrcr}
\hline
\hline
Object & TIC &  Start time & \multicolumn{1}{c}{\textit{N}} & \multicolumn{1}{c}{$\delta T$} & \textit{TESS} mag & Sect. & \multicolumn{1}{c}{CROWDSAP} & \multicolumn{1}{c}{$0.1\%$\,FAP} \\
& & (BJD-2\,457\,000) & & \multicolumn{1}{c}{(d)} & & & & \multicolumn{1}{c}{(mma)}\\
\hline
MCT~2148-2911 & 053851007 & 1325.301 & 18\,094 & 27.9 & 16.1 & 1 & 0.17 & 6.19 \\
HE~0031-5525 & 281594636 & 1354.113 & 18\,313 & 27.4 & 15.8 & 2 & 0.28 & 5.47 \\
EC~00497-4723 & 101916028 & 1354.113 & 18\,315 & 27.4 & 16.5 & 2 & 0.85 & 8.35 \\
MCT~0016-2553 & 246821917 & 1354.114 & 18\,317 & 27.4 & 15.9 & 2 & 0.13 & 4.92 \\ 
WD~0108-001 & 336891566 & 1385.954 & 13\,451 & 20.3 & 17.0 & 3 & 0.92 & 10.93 \\
HS~0235+0655 & 365247111 & 1410.908 & 15\,751 & 25.9 & 16.5 & 4 & 0.89 & 12.42 \\
KUV~03442+0719 & 468887063 & 1437.997 & 17\,631 & 26.0 & 16.6 & 5 & 0.57 & 16.62 \\
WD~J0925+0509 & 290653324 & 1517.403 & 13\,384 & 24.6 & 15.3 & 8 & 0.56 & 3.81 \\
HS~1013+0321 & 277747736 & 1517.402 & 13\,390 & 24.6 & 15.7 & 8 & 0.97 & 4.59 \\ 
EC~11266-2217 & 219442838 & 1544.278 & 15\,569 & 24.2 & 16.4 & 9 & 0.35 & 5.20 \\ 
\hline
\end{tabular}
\end{table*}

We downloaded the light curves from the \textit{Mikulski Archive for Space Telescopes} (MAST), and extracted the PDCSAP fluxes provided by the Pre-search Data Conditioning Pipeline \citep{2016SPIE.9913E..3EJ} from the fits files. We removed all data points with quality warning flags, and finally we corrected the light curves for long-period systematics. We normalised the light curves by fitting a fourth-order Savitzky-Golay filter with a three-day window length computed with the Python package \textsc{lightkurve} \citep{2019AAS...23310908B}. This correction does not affect the frequency domain of the short-period white dwarf pulsations. The panels of Figs.~\ref{fig:lcvar} and \ref{fig:lcnov} show the resulting light curves. Note that the Pre-search Data Conditioning pipeline corrects the flux for
each target to account for crowding from other stars.

Fig.~\ref{fig:lcvar} shows the stars that are confirmed to vary, while Fig.~\ref{fig:lcnov} show those that do not show variability in \textit{TESS}. Note that \ec, \hs, \ross, \bpmb, \mct, and \lnt\ were observed in one sector, \bpm\ was measured in two, while \he, located in the \textit{TESS} Continuous Viewing Zone (CVZ), was observed in 13 consecutive sectors.
Tables~\ref{tabl:journal1} and \ref{tabl:journal2} show the journal of observations of our targets observed and not observed to vary by \textit{TESS}, respectively.

\section{Light-curve analysis}
\label{sect:analyses}

We performed standard Fourier analysis and pre-whitening of the data sets with the photometry modules of the Frequency Analysis and Mode Identification for Asteroseismology (\textsc{famias}) software package \citep{2008CoAst.155...17Z}\footnote{\textsc{famias} is a package of software tools for the analysis of photometric and spectroscopic time-series data. It enables us to search for periodicities in these data sets using the method of Fourier analysis and non-linear least-squares fitting techniques. \textsc{famias} is also capable of performing mode identification of the detected periodicities utilising different photometric and spectroscopic methods.}. We also used in-house developed software for the least-squares fitting, used for the analysis of Whole Earth Telescope \citep{1990ApJ...361..309N}, \textit{Kepler}, and \textit{K2} data. 
We chose the detection limit to be at $0.1$ per cent false alarm probability (FAP), that is, in this case, there is $99.9$ per cent chance that a peak reaching this limit is not just result of noise fluctuations. 

We calculated the $0.1\%$ FAP threshold following the method described in \citet{2016A&A...585A..22Z}: we generated 10\,000 synthetic light curves of Gaussian random noise for the times of the real observations, calculated their Fourier transforms up to the Nyquist limit ($\approx4167\,\mu$Hz), and then determined the probability for the different signal-to-noise (S/N) ratios that a single peak emerges at that S/N value only due to noise fluctuations. The noise level was chosen to be the mean amplitude level of the given synthetic light curve. This way we could calculate the S/N values corresponding to the $0.1\%$ FAP thresholds, and we accepted a frequency peak only above this S/N limit as significant.

However, in some cases, we were unable to remove all of the observed power by simple pre-whitening, suggesting the presence of amplitude/phase variations in the signal. We fitted Lorentzian envelopes to these peaks as described in \citet{2015ApJ...809...14B} and \citet{2017ApJS..232...23H}. We fitted the peaks in the power spectrum with the function presented in eq.~1 of \citet{2015ApJ...809...14B} and eq.~1 of \citet{2017ApJS..232...23H}. The Fourier transforms were oversampled by a factor of six.


\subsection{\ross}\label{sec:ross}

\subsubsection{Ground-based observations}
\ross\, (TIC~29854433, $G=14.23$\,mag\footnote{\textit{Gaia} Data Release 2 $G$ magnitude, \citet{2018yCat.1345....0G}}, $\alpha_{2000}=01^{\mathrm h}36^{\mathrm m}14^{\mathrm s}$, $\delta_{2000}=-11^{\mathrm d}20^{\mathrm m}33^{\mathrm s}$) is also known as ZZ~Ceti, and is the namesake of the class of hydrogen-atmosphere pulsating white dwarfs. It was one of the first pulsating white dwarfs discovered by \citet{1971ApJ...163L..89L}. The star has shown a remarkably stable pulsation spectrum for over 48 years, with two dominant periodicities (each now known to be triplet ($\ell = 1$) modes). \citet{1971ApJ...163L..89L} isolated two modes with periods of 212.864 s (4697.8 $\mu$Hz) and 273.0 s (3663 $\mu$Hz). These relatively short periods are consistent with its position near the blue edge of the ZZ Ceti instability strip, with an effective temperature of 12\,200\,K \citep[and references therein]{2015ApJ...815...56G}.

The stability of the pulsation amplitudes and frequencies has been studied in depth, in part to determine the secular period change rates driven by the cooling of the star.  This decades--long effort is summarised in, for example, \citet{2013ApJ...771...17M}.  The two main modes identified by \citet{1971ApJ...163L..89L} have been resolved as triplets, with the amplitude of the central $m=0$ peak about 3-5 times smaller than  the flanking $m=\pm 1$ peaks.  The splitting is 4.0~$\mu$Hz for the 213\,s mode, and 3.5~$\mu$Hz for the 275~s mode.

\citet{2013ApJ...771...17M} report that the amplitudes of the $m=\pm 1$ modes in \ross\,  have been very stable -- a fact that has enabled measurement of the long--term secular change in the pulsation phase, and therefore the rate of period change.  This decades-long stability in principle allows us to use \ross\, to help calibrate the pulsation amplitudes of white dwarfs observed by {\it TESS} without having simultaneous ground-based observations.

In Table~\ref{tabl:rossgb}, we present the weighted average amplitudes of the largest-amplitude modes in \ross. Amplitudes we used are those reported by \citet{2015ApJ...815...56G}, \citet{2013ApJ...771...17M}, and references therein.

\begin{table*}
\centering
\caption{Periods and frequencies derived by ground-based observations on \ross. The next-to-last column lists the corresponding weighted mean amplitudes (near $V$ magnitude), while the last column is the expected amplitude in the {\it TESS} bandpass assuming a blackbody spectral energy distribution as per Eq.~\ref{eg:ross}.}
\label{tabl:rossgb}
\begin{tabular}{ccccc}
\hline
\hline
  & Period [s] & Frequency [$\mu$Hz] & Amplitude [mma] & Amplitude at 825\,nm \\
\hline
$f_1$ & 212.8 & 4699.2 & 4.28(4) & 2.71 \\
$f_2$ & 213.1 & 4692.6 & 6.86(4) & 4.34 \\
$f_3$ & 274.3 & 3646.3 & 4.41(4) & 2.79 \\
$f_4$  & 274.8 & 3639.4 & 3.17(4) & 2.01 \\
\hline
\end{tabular}
\end{table*}

The bandpass of {\it TESS} is much redder than that of ground based photomultiplier and CCD observations, but we can compute the expected amplitude in the {\it TESS} bandpass quite readily.  Since the $g-$mode pulsations reflect temperature changes across the white dwarf surface, we can (assuming a blackbody with $T \approx$ 12\,200\,K) compute the amplitude ratio in the {\it TESS} bandbpass (centred at approximately 825\,nm) to that in the optical (assumed to be 425\,nm), following \citet{1994AJ....107..298K}:
\begin{equation}
\label{eg:ross}
    \frac{A_{\rm 825}}{A_{\rm 425}} = \frac{425}{825} \ \  e^{-16450/T} \ \  \frac{e^{33890/T}-1}{e^{17458/T}-1}
\end{equation}
For $T_{\rm eff}$=12\,200\,K, this implies that the amplitudes observed by {\it TESS} should be about 63\% of those observed in the optical. The last column in Table~\ref{tabl:rossgb} are therefore the expected amplitudes for {\it TESS} observations.
We compiled the published observed amplitudes of \ross\, to obtain an estimate of the expected amplitude for our observations with {\it TESS}.

\subsubsection{\textit{TESS} observations}

{\it TESS} observed \ross\, during Sector 3.  The light curve is shown in Fig.~\ref{fig:lcvar}.  Note that towards the end of the first segment, the noise level in the light curve increased (a common issue for Sector 3 data). We chose to retain those data to improve the duty cycle after examining the periodogram with and without the problematic segment to ensure that the noise did not adversely impact the frequency analysis.  

\begin{figure*}
\centering
\includegraphics[width=18cm]{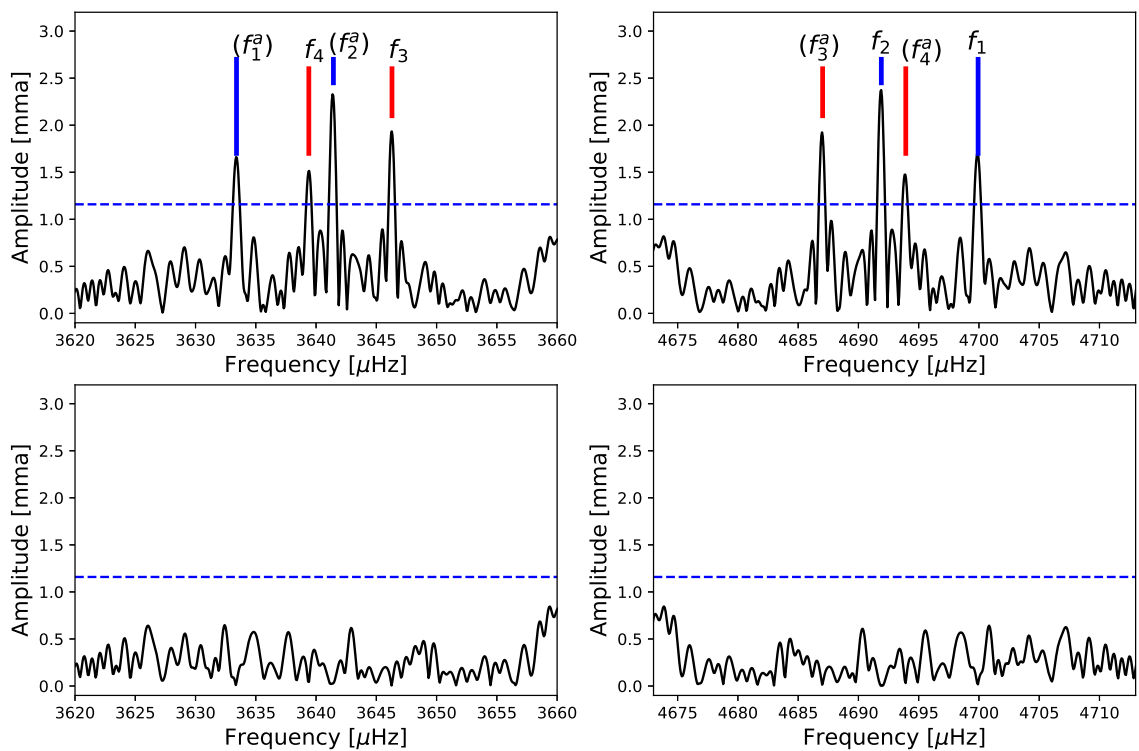}
\caption{Fourier transform of the {\it TESS} light curve of \ross. The panels on the left show the sub-Nyquist frequency range where the signal exceeds the 0.1\% FAP (S/N=4.54, horizontal dashed line). The peaks correspond to  $f_3$ and $f_4$ from Table~\ref{tabl:rossgb} (red lines), and the aliases (reflected across the Nyquist frequency) of $f_1$ and $f_2$ (blue vertical lines).  The right-hand panels show the transform above the Nyquist frequency, where the real periodicity -- alias situation is reversed.  The two bottom panels show the residuals in the spectrum after pre-whitening by the least-squares fit sinusoids $f_1$, $f_2$, $f_3$, $f_4$ listed in Table~\ref{tabl:rosslsq}.}
\label{fig:rossFT-z}
\end{figure*}

Our 0.1\% FAP level is at a S/N of 4.59 (corresponding to an amplitude of 1.24 mma).  In the periodogram, the only peaks that rose above that threshold were precisely at the frequencies corresponding to the known pulsation modes (or their Nyquist aliases) listed in Table~\ref{tabl:rossgb}.  While the 275\,s modes (at 3646.3 and 3639.4\,$\mu$Hz) were below the Nyquist frequency of 4167~$\mu$Hz, the 213\,s modes (at 4692.6 and 4699.2 $\mu$Hz) are above the Nyquist frequency.  They are ``reflected'' back across the Nyquist frequency to 3640.7 and 3634.1\,$\mu$Hz.  As we show in Fig.~\ref{fig:rossFT-z}, the two aliases of the 213\,s modes intermingle with the two 275\,s peaks.  Beyond the Nyquist frequency, the same pattern repeats, with the alias of the 275\,s modes intermingling with the ``true'' 213\,s peaks.  We performed a least-squares fit to the time series of \ross\ using the prior values of the four frequencies as a starting point.  The formal fit to the light curve is given in Table~\ref{tabl:rosslsq}.  Pre-whitening the data using these four frequencies completely removes all peaks in this region, and their mirror aliases, at or near the 0.1\% FAP.

For more details considering the super-Nyquist analyses of compact pulsators, see the papers of \citet{2012MNRAS.424.2686B} and \citet{2017ApJ...851...24B}.

Finally, for comparison with the ground-based measurements of the amplitudes of the modes of \ross, we need to correct the fit values of the amplitudes for the fact that the frequencies lie close to (and beyond) the Nyquist frequency.  Thus the signals are attenuated by the smoothing/averaging over the 120\,s {\it TESS} integrations.  Assuming a sinusoidal waveform, the functional form of the attenuation is a sinc function in frequency, with the first zero at twice the Nyquist frequency.  The corrected amplitudes are in the fifth column of Table~\ref{tabl:rosslsq}. The corrections were applied to the uncertainties, and the ground-based optical amplitudes were reduced by the bandpass and integration near the Nyquist limit to compare with the \textit{TESS} observed values.

Considering the 11-frequency solution presented by \citet{2015ApJ...815...56G}, we found that only the four highest-amplitude modes can be seen in the \textit{TESS} data set. The rest, with ground-based amplitudes of 0.3--1.4\,mma, all fall below the detection limit.

\begin{table*}
\centering
\caption{Weighted mean amplitudes (from ground-based photometry) for the dominant pulsation modes in \ross. The fifth column shows the amplitude expected in the {\it TESS} bandpass, corrected for amplitude dilution near the Nyquist frequency.}
\label{tabl:rosslsq}
\begin{tabular}{cccccc}
\hline
\hline
  & Period [s] & Frequency [$\mu$Hz] & amp [mma] & {\it TESS} pred [mma] & {\it TESS} obs [mma] \\
\hline
$f_1$ & 212.7704(17) & 4699.900(37) & 4.28(4) & 1.82(2) & 1.3(2) \\
$f_2$ & 213.1348(12) & 4692.868(26) & 6.86(4) & 2.93(2) & 2.4(2) \\
$f_3$ & 274.2521(24) & 3646.280(32) & 4.41(4) & 2.42(2) & 1.9(2) \\
$f_4$  & 274.7719(28) & 3639.383(37) & 3.17(4) & 1.74(2) & 1.8(2) \\
\hline
\end{tabular}
\end{table*}

Comparing the (corrected) observed amplitudes from {\it TESS} to the expected values from ground-based data over the past 48 years, we see that there is close agreement for all 4 modes resolved by {\it TESS}.  Therefore, we expect that amplitudes from {\it TESS} can be directly compared with archival and future ground-based measurement of the amplitude of white dwarf pulsations (after correcting for different bandpasses and exposure times).  


\subsection{\ec}

\subsubsection{Ground based observations}

The light variations of \ec\ (TIC\,033986466, $G=15.38$\,mag, $\alpha_{2000}=23^{\mathrm h}51^{\mathrm m}22^{\mathrm s}$, $\delta_{2000}=-24^{\mathrm d}08^{\mathrm m}17^{\mathrm s}$) were discovered by \citet{1993MNRAS.263L..13S}. They revealed a complex frequency structure by their six-night measurements. They determined four pulsation frequencies at 1007.07, 1243.07, 1151.79, and 1010.86\,$\mu$Hz (992.98, 804.46, 868.21, and 989.26\,s, respectively). They also detected the second harmonics of several frequencies, and additional closely spaced frequency components at the 1010.86\,$\mu$Hz peak. Apart from the discovery paper, the results of only one observation have been published up to now: \citet{2005ASPC..334..471T} presented a two-frequency solution with periods at 878.8 and 508.1\,s, based on time-series spectroscopy.

\subsubsection{Results of the \textit{TESS} observations}

\begin{figure*}
\centering
\includegraphics[width=18cm]{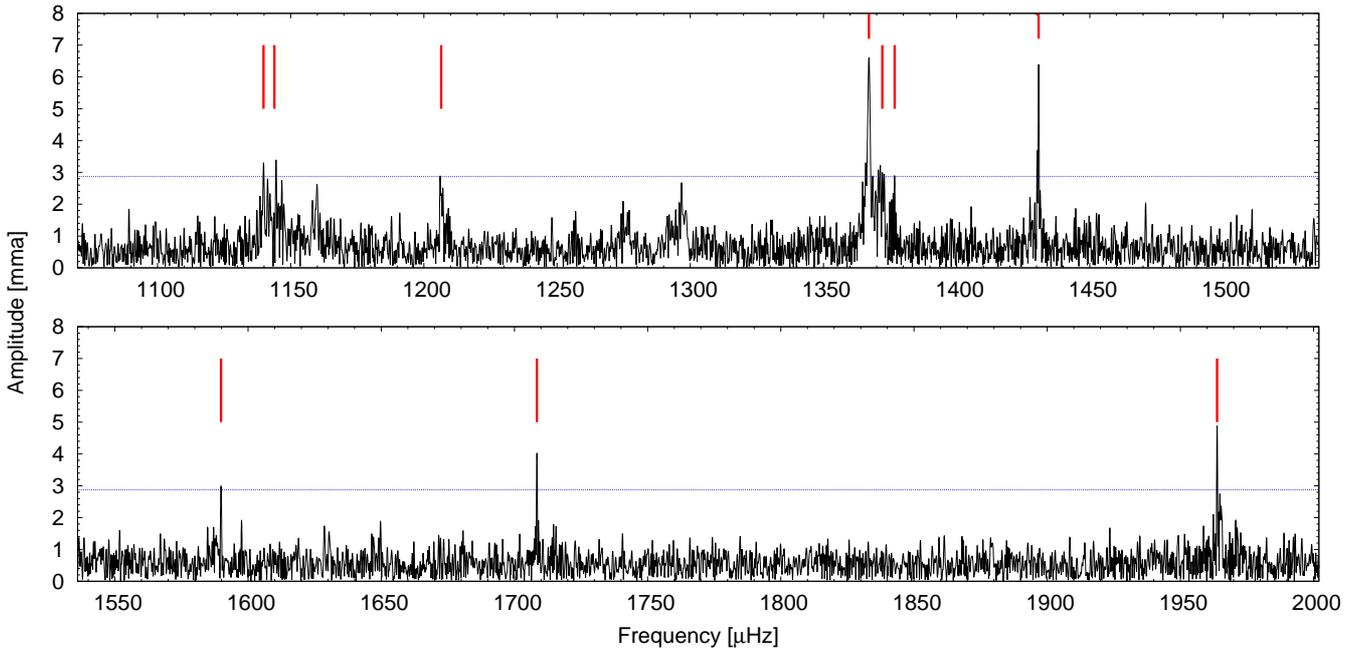}
\caption{Fourier transform of the light curve of \ec. Red lines mark the frequencies listed in Table~\ref{tabl:ecfreq2}, while the blue dashed line represents the $0.1\%$ FAP level (S/N=4.74) calculated by this Fourier transform.}
\label{fig:ecfreq}
\end{figure*}

\begin{table*}
\centering
\caption{This work: results of the fit of the \ec\ data set with 10 frequencies. The errors are formal uncertainties. The results of Lorentzian fits are marked with star symbols ($^*$). In the case of the supposed rotationally split frequencies, we mark the possible azimuthal order values. For completeness, we also list the periods detected by \citet{1993MNRAS.263L..13S} and \citet{2005ASPC..334..471T}.}
\label{tabl:ecfreq2}
\begin{tabular}{llclcc}
\hline
\hline
 & \multicolumn{3}{c}{This work} & \multicolumn{1}{c}{\citet{1993MNRAS.263L..13S}} & \multicolumn{1}{c}{\citet{2005ASPC..334..471T}} \\
 & \multicolumn{1}{c}{Frequency} & \multicolumn{1}{c}{Period} & \multicolumn{1}{c}{Ampl.} & \multicolumn{1}{c}{Period} &  \multicolumn{1}{c}{Period} \\
 & \multicolumn{1}{c}{[$\mu$Hz]} & \multicolumn{1}{c}{[s]} & \multicolumn{1}{c}{[mma]} & \multicolumn{1}{c}{[s]} &  \multicolumn{1}{c}{[s]}\\
\hline
 & & & & 993.0 & \\
 & & & & 989.3 & \\
$f_{1}^{-1/0}$ & 1139.799(34) & 877.348 & 3.30(49) &  & 878.8 \\
$f_{1}^{0/+1}$ & 1144.67(35)$^*$ & 873.614 & 1.61(30) & & \\
 & & & & 868.2 & \\
$f_{2}$ & 1206.495(39)$^*$ &	828.847 & 2.41(05) & & \\
 & & & & 804.5 & \\
$f_{3}^{-1}$ & 1367.082(17) &	731.485	& 6.65(49) & & \\
$f_{3}^{0}$ & 1372.116(35) &	728.802	& 3.15(49) & & \\
$f_{3}^{+1}$ & 1376.720(39) &	726.364	& 2.81(49) & & \\
$f_{4}$ & 1430.780(17) &	698.920	& 6.41(49) & & \\
$f_{5}$ & 1589.850(37) &	628.990 & 2.96(49) & & \\
$f_{6}$ & 1708.434(27) &	585.331 & 4.05(49) & & \\
$f_{7}$ & 1963.783(23) &	509.221	& 4.88(49) & & 508.1 \\
\hline
\end{tabular}
\end{table*}

If we pre-whiten the data set with the standard procedure down to the significance threshold, we can identify 17 frequencies above the $0.1\%$ FAP limit (S/N\,=\,4.74). However, these are not all independent pulsation modes, but they form groups of peaks in some cases, suggesting amplitude/phase variations occurring during the \textit{TESS} observations. Actually, we found that the high frequency peaks (those with periods shorter than 800\,s) are stable enough in amplitude and phase that pre-whitening leaves no residuals above the detection threshold. However, significant residuals remain after removing the lower frequency peaks.

This feature is consistent with fig.~5 of \citet{2017ApJS..232...23H}, that shows the stochastic nature appearing at periods longer than about 800\,s. 

\begin{figure*}
\centering
\includegraphics[width=18cm]{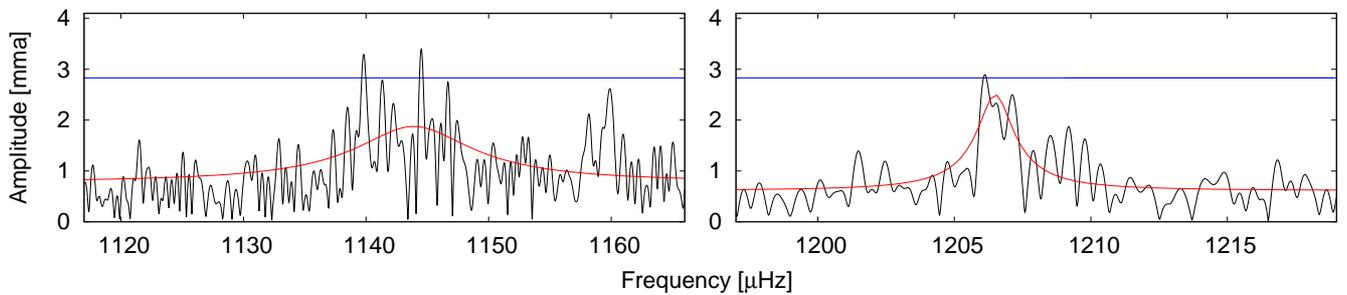}
\caption{Lorentzian fits (red lines) to the frequency regions at 1143 and 1206\,$\mu$Hz. Blue lines denote the 0.1\% FAP level.}
\label{fig:ecfit}
\end{figure*}

Therefore, we fit a Lorentzian envelope to the regions at 1144 and 1206\,$\mu$Hz (see Fig.~\ref{fig:ecfit}). The dominant frequency at $1367\,\mu$Hz ($\sim 730$\,s) shows intermediate behaviour -- removing the main peak leaves behind some additional peaks, but that one peak's dominance makes fitting a Lorentzian challenging. The damping times calculated for the 1144 and 1206\,$\mu$Hz regions are 0.183 and 2.252 days with Lorentzian widths of $25.5 \pm 1.0\,\mu$Hz and $1.59 \pm 0.11\,\mu$Hz, respectively.

Table~\ref{tabl:ecfreq2} lists the results of the 10-frequency fit, including the parameters of the Lorentzian fits, while Fig.~\ref{fig:ecfreq} shows the Fourier transform of the whole light curve. Red dashed lines mark the frequencies listed in Table~\ref{tabl:ecfreq2}.

Note that the dominant peak and the peaks at 1372.116 and 1376.720\,$\mu$Hz form a triplet with 5.034 and 4.604\,$\mu$Hz frequency separations, respectively, while we see similar frequency difference of 4.085\,$\mu$Hz between the peaks at 1143.895 and 1139.810\,$\mu$Hz. This suggests that we see the results of rotational frequency splitting at least in the first case. Considering the 4.085\,$\mu$Hz separation, the approximate rotation period of the star could be 1.42 or 2.36\,d, assuming the rotational splitting of $\ell=1$ or $\ell=2$ modes, respectively.  


We also mark in Table~\ref{tabl:ecfreq2} the frequencies obtained by ground-based observations. Apparently, \citet{1993MNRAS.263L..13S} detected a set of modes completely different from those we determined by the \textit{TESS} observations, as if we observed two different stars. Besides the detected complex frequency structure with closely spaced peaks in the Fourier transforms, this also suggests that some of the mode amplitudes are not stable. 


\subsection{\bpm}

\subsubsection{Ground-based observations}

\bpm\ (TIC\,101014997, $G=15.06$\,mag, $\alpha_{2000}=03^{\mathrm h}43^{\mathrm m}29^{\mathrm s}$, $\delta_{2000}=-45^{\mathrm d}49^{\mathrm m}04^{\mathrm s}$) was discovered as the 10th member of the ZZ~Ceti class by \citet{1976ApJ...210L..35M}. The authors presented the results of four nights of observations. They reported an interesting change in the pulsational behaviour of \bpm. On the discovery night it showed light variations with 314\,s, but on the subsequent nights, its periodicity changed to 617\,s. However, it turned out later, that this change was not real, but only result of erroneous data reduction of the first night's measurements \citep{1986MNRAS.220P..19O}. \citet{1976ApJ...210L..35M} also detected the two harmonics of the 617\,s mode ($f_0$), significant power in the spectra near (3/2)$f_0$, and combinations of frequencies.

\citet{1992MNRAS.258..415O} presented the results of the long-term investigations of \bpm\, carried out between 1975 and 1989. They found the dominant frequency to be at $f_0 = 1.62$\,mHz (617\,s) again, and detected significant peaks at multiples of $1.54 f_0$. In some cases, further frequencies appeared at multiples of $1.48 f_0$. They also identified, that at $f_0$ there is actually a rotationally split triplet with $12.7\,\mu$Hz frequency separations. Besides these, they detected the harmonics of $f_0$, an additional frequency at 1.8778\,mHz (533\,s), and several combination terms. Considering the stability of frequencies, they found variability in the period of $f_0$, and also variable amplitudes at the other frequencies from season to season. The authors summarised their findings as the observed structure of the frequencies of \bpm\ can be described by only a few independently excited modes, which drive others through non-linear coupling by direct resonance.

\subsubsection{\textit{TESS} observations}

The dominant peak is at $1618.405\,\mu$Hz, in agreement with the results of the previous observations. 
This peak is also the central component of a triplet with $13.42\,\mu$Hz frequency separations. This is also in good agreement with the $12.7\,\mu$Hz finding of \citet{1992MNRAS.258..415O} by ground-based observations. Note that there are also residual closely spaced peaks at the triplet components. These are partly caused by the (apparent) changing noise level between the two sectors and sub-sector parts. 

We also detected the second harmonic of the dominant frequency, significant peaks at 1.54 and 2.54 times the dominant frequency, two independent frequencies at $1461.119$ and $1548.455\,\mu$Hz, a combination term at $3166.876\,\mu$Hz, and a low-frequency peak corresponding to the frequency separation of the triplet components at $13.409\,\mu$Hz. 

Besides these, four additional peaks can be determined by the data set, close to, but well-separated from the peaks at 1.54, 2, and 2.54 times the dominant peak, respectively. These are at $f_w=2483.202$, $f_x=3245.652$, $f_y=3252.593$, and $f_z=4115.083\,\mu$Hz. It is remarkable, that the spacing between $f_w$ and $1.54f_1$ is $6.952\,\mu$Hz, which is nearly half the spacing within the triplet around the dominant peak. This is also true for the spacings between $f_x$ and $f_y$ ($\delta f=6.941\,\mu$Hz), and $f_z$ and $2.54f_1$ ($\delta f=6.50\,\mu$Hz). The origin of these additional peaks is in question.

\begin{figure*}
\centering
\includegraphics[width=18cm]{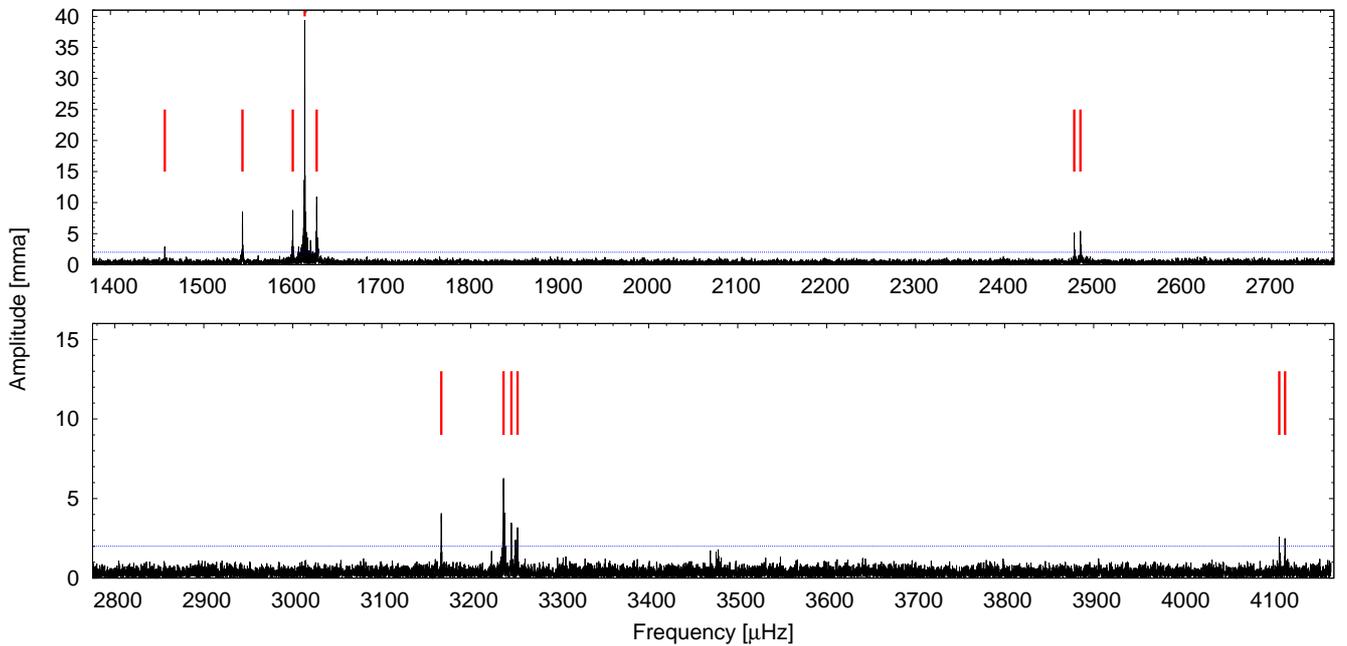}
\caption{Fourier transform of the light curve of \bpm. Red lines mark the frequencies listed in Table~\ref{tabl:bpmfreq2}, while the blue dashed line represents the $0.1\%$ FAP level (S/N=4.85) calculated by this Fourier transform.}
\label{fig:bpmfreq}
\end{figure*}

Table~\ref{tabl:bpmfreq2} lists the 14-frequency solution without the closely spaced residual peaks (the $0.1\%$\,FAP limit is S/N=4.85). These frequencies are also marked in Fig.~\ref{fig:bpmfreq}, showing the Fourier transform of the whole light curve.

\begin{table*}
\centering
\caption{14-frequency solution of the \bpm\ data set. The errors are formal uncertainties calculated by \textsc{famias}. In the case of the rotationally split frequency, we mark the azimuthal order values.}
\label{tabl:bpmfreq2}
\begin{tabular}{lrrr}
\hline
\hline
 & \multicolumn{1}{c}{Frequency [$\mu$Hz]} & \multicolumn{1}{c}{Period [s]} & \multicolumn{1}{c}{Amplitude [mma]} \\
\hline
$f_3-f_3^{\pm 1}$ & 13.409(12) & 	74577 & 	3.08(32) \\
$f_1$ & 1461.117(13) & 	684.408 & 	2.87(32) \\
$f_2$ & 1548.455(4) & 	645.805 & 	8.47(32) \\
$f_3^{-1}$ & 1604.982(4) & 	623.060 & 	8.94(32) \\
$f_3$ & 1618.401(1) & 	617.894 & 	39.37(32) \\
$f_3^{+1}$ & 1631.844(4) & 	612.804 & 	10.80(32) \\
$f_w$ & 2483.203(7) & 402.706 & 5.06(32) \\
$1.54f_3$ & 2490.147(7) & 	401.582 & 	5.35(32) \\
$f_2 + f_3$ & 3166.877(9) & 	315.769 & 	4.05(32) \\
$2f_3$ & 3236.794(6) & 	308.948 & 	6.26(32) \\
$f_x$ & 3245.648(11) & 308.105 & 3.41(32) \\
$f_y$ & 3252.590(12) & 307.447 & 3.16(32) \\
$2.54f_3$ & 4108.584(15) & 	243.393 & 	2.49(32) \\
$f_z$ & 4115.083(16) & 243.009 & 2.41(32) \\
\hline
\end{tabular}
\end{table*}

\subsection{\bpmb}

\subsubsection{Ground-based observations}
\bpmb\ (TIC\,102048288, $G=15.48$\,mag, $\alpha_{2000}=01^{\mathrm h}06^{\mathrm m}54^{\mathrm s}$, $\delta_{2000}=-46^{\mathrm d}08^{\mathrm m}54^{\mathrm s}$) was one of the first ZZ Ceti pulsators discovered; the first observations are described by \citet{1976ApJ...209..853H}. They identified several periodicities between 1190 and 1640\,$\mu$Hz in 24 observing runs spanning several months. The power spectrum was variable, but most of the peaks in the Fourier transforms of individual runs were situated in the 1000\,--\,1900\,$\mu$Hz domain (see Table~\ref{tabl:bpmbfreq}).  In retrospect, we recognise this as typical behaviour of cool DAV stars, with complex pulsation spectra of modes with periods longer than 800 seconds \citep{2017ApJS..232...23H}. To our knowledge, no time-series photometric observations of \bpmb\ have been performed since the discovery observations in 1974. Given this, the low-resolution Fourier transforms available in \citet{1976ApJ...209..853H} make it difficult to compare with the \textit{TESS} results.

\begin{table*}
\centering
\caption{Consistent periodicities in the ground-based observations of \bpmb\ determined by \citet{1976ApJ...209..853H}.}
\label{tabl:bpmbfreq}
\begin{tabular}{lrrr}
\hline
\hline
 & \multicolumn{1}{c}{Frequency [$\mu$Hz]} & \multicolumn{1}{c}{Period [s]} & \multicolumn{1}{c}{Amplitude range [mma]} \\
\hline
$f_1$      & 1643.1 & 608.6 & 12--15 \\
$f_2$ & 1464.8 & 682.7 & 10--12 \\
$f_3$ & 1342.8 & 744.7 & 7 \\
$f_4$ & 1190.2 & 840.2 & 15 \\
$f_5$  & 1159.7 & 862.3 & 6--15 \\
\hline
\end{tabular}
\end{table*}

\subsubsection{\textit{TESS} observations}
\bpmb\ was observed for 25.4 days during the 27.5 day observing sector 2, so therefore with a duty cycle of 92.5\%. The light curve of \bpmb\ (see Fig.~\ref{fig:lcvar}) shows no evidence for bursts, which occur in some DA white dwarfs that pulsate in similar periods \citep{2016ApJ...829...82B}.  There were no apparent low-frequency variations (1 day or longer) following examination of the un-flattened light curve.

\begin{figure*}
\centering
\includegraphics[width=18cm]{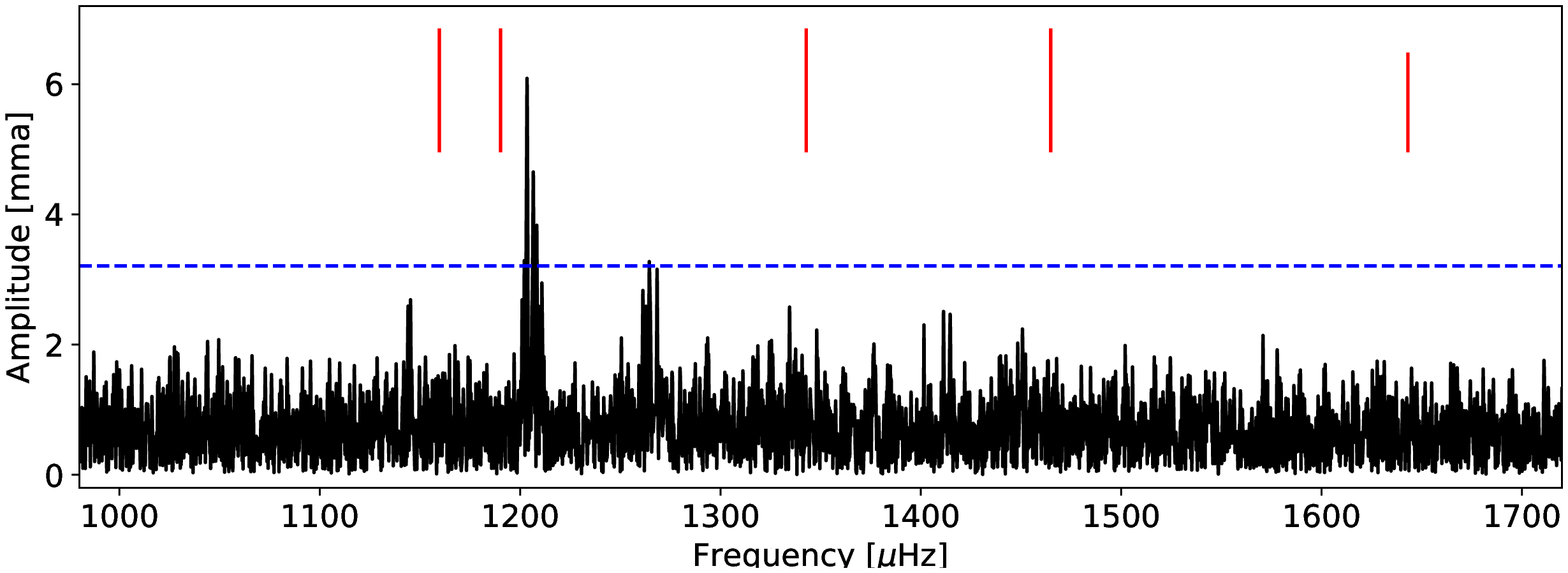}
\caption{Fourier transform of the light curve of \bpmb\ in the only spectral region where signals reached above the 0.1\%FAP level (S/N=4.69, blue dashed line). Red lines mark the approximate frequencies seen in the 1975 data reported in \citet{1976ApJ...209..853H} and listed in Table~\ref{tabl:bpmbfreq}.}
\label{fig:bpmbFT}
\end{figure*}

The Fourier transform of the light curve of \bpmb\ -- the first observations since the mid 1970s -- confirms that it is indeed a DAV pulsator, with the largest-amplitude variations at frequencies between 1100 and 1300\,$\mu$Hz.  Fig.~\ref{fig:bpmbFT} shows the FT over the range of frequencies reported by \citet{1976ApJ...209..853H}, with red lines showing the approximate positions of their identifications.  In the \textit{TESS} data, the only significant signals (i.e. S/N above 4.69) are in a narrow frequency range between 1195 and 1215\,$\mu$Hz, with a second region around 1260\,$\mu$Hz that lies just above the significance threshold.

\begin{figure*}
\centering
\includegraphics[width=18cm]{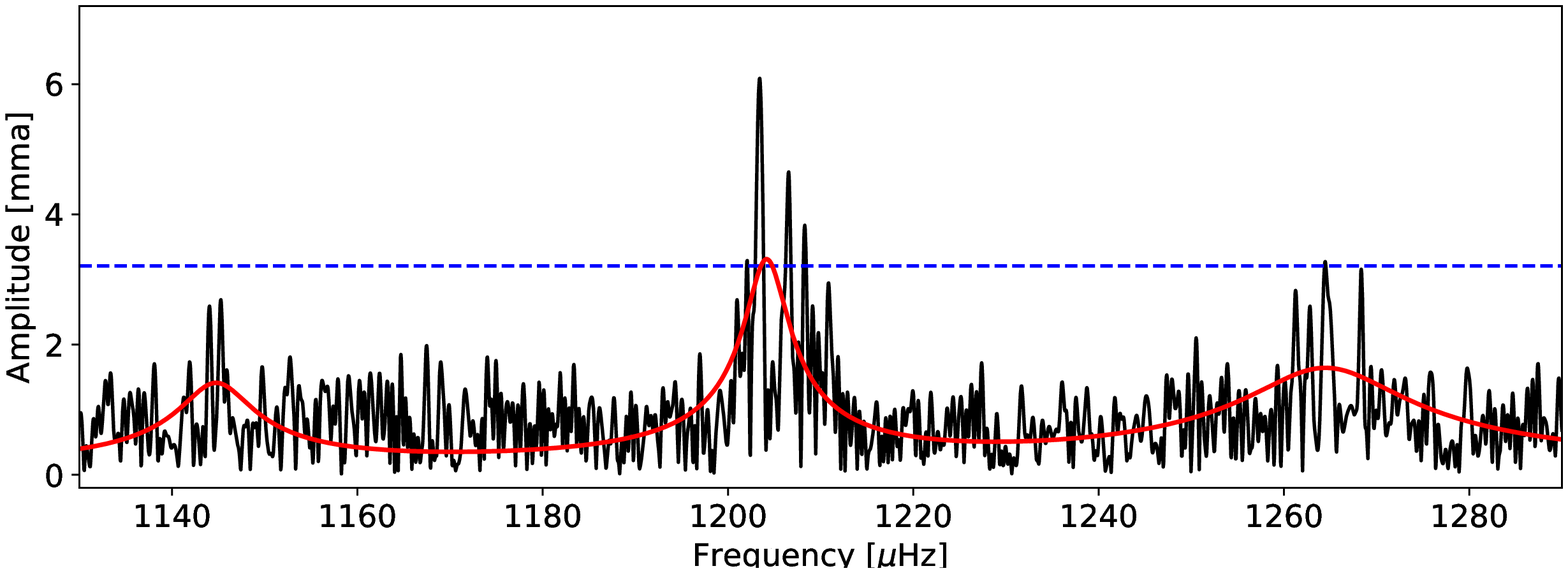}
\caption{Area of maximum signal in the Fourier transform of \bpmb. The region of largest signal is complex. The red curve is a fit of Lorentzian envelopes to the main signal, and the two flanking regions that do not reach the significance threshold.}
\label{fig:bpmbFTz}
\end{figure*}

An expanded view of the FT of the \bpmb\ data is shown in Fig.~\ref{fig:bpmbFTz}. The signal in the range from 1200--1215\,$\mu$Hz is extremely complex, indicating that the mode or modes of pulsation are not coherent over the 27 day duration of the run.  In fact, this is very similar to what is seen in the longer-period pulsations of DAV stars near the cool end of the instability strip, which is reminiscent of stochastically driven oscillations seen in solar-type asterosesimic data. Attempts to identify and remove coherent periodicities in this region were unsuccessful -- pre-whitening was unable to remove all of the observed power, and as individual coherent periodicities were removed, new nearby peaks appeared. Therefore, we fit a Lorentzian envelope to this region, and the region around 1260\,$\mu$Hz. The central frequencies of the Lorentzian fits to the frequency groups are as follows: $1204.17 \pm 0.02\,\mu$Hz, and $1264.49 \pm 0.16\,\mu$Hz. Note that as Fig.~\ref{fig:bpmbFTz} shows, we can fit a third frequency group at $\sim 1140\,\mu$Hz with the central frequency at $1144.75 \pm 0.15\,\mu$Hz, even though the corresponding peaks are below our detection threshold. 

We find a damping time for the strongest modes of about 0.77 days; thus the 27 day observing span dilutes the true power of the oscillations over a broad range of frequencies.  Shorter observations, such as a typical 1 night terrestrial observation which is shorter than the typical damping time for these modes, would be expected to see much large amplitude pulsations.  Indeed, simulations show that modes which conform to the fitted Lorentzian power envelope widths and heights can produce peaks in excess of 20\,mma within that frequency range.  Thus the results from \citet{1976ApJ...209..853H} are consistent with what \textit{TESS} observed.  However, the modes seen by \citet{1976ApJ...209..853H} at other frequencies are (apparently) either too low in amplitude to be detected, or are no longer present.

\begin{table*}
\centering
\caption{Results of the Lorentzian fits to the frequency groups observed in the \bpmb\ data set.  The central frequencies are marked with star symbols ($^*$), as in the case of \ec\ (Table~\ref{tabl:ecfreq2}). We also listed the Lorentzian widths for the different fits.}
\label{tabl:bpmbfreq2}
\begin{tabular}{lrrrr}
\hline
\hline
 & \multicolumn{1}{c}{Frequency [$\mu$Hz]} & \multicolumn{1}{c}{Period [s]} & \multicolumn{1}{c}{Amplitude [mma]} & \multicolumn{1}{c}{Width [$\mu$Hz]}\\
\hline
$f_1$ & 1144.760(150)$^*$ & 873.546 & 1.39(03) & 8.30(43) \\
$f_2$ & 1204.487(021)$^*$ & 830.229 & 3.14(02) & 4.41(06) \\
$f_3$ & 1264.164(093)$^*$ & 789.163 & 1.83(02) & 9.59(26) \\
\hline
\end{tabular}
\end{table*}


\subsection{\mct}

\subsubsection{Ground-based observations}
The pulsating behaviour of the DAV \mct\ (TIC~164772507, $G=15.23$\,mag, $\alpha_{2000}=01^{\mathrm h}47^{\mathrm m}22^{\mathrm s}$, $\delta_{2000}=-21^{\mathrm d}56^{\mathrm m}51^{\mathrm s}$) was announced concurrently with \he\ in \citet{2003ApJ...591.1184F}. It was observed over two consecutive nights for 3740~s on the first and 5780~s on the second, each with sampling times of 10\,s. They identified three periodicities at 1215, 1374, and 2164\,$\mu$Hz with amplitudes ranging from 17 to 25\,mma; they noted that they expected that more periods could likely be resolved with longer observations. They determined the temperature of the star to be around 11\,550\,K, making this another example of a DAV being in the middle of the instability strip.

These observations were followed up by \citet{2014MNRAS.437.1836K}. They observed for 50\,min with a sampling time of 0.1\,s. The frequency resolution of their data was approximately 8\,$\mu$Hz, estimated using their data using Equation 5.52 of \citet{2010aste.book.....A}.  They identified three periodicities in this short data set at 1370, 2200, and 2290\,$\mu$Hz with amplitudes between 6 and 10\,mma which, given the brevity of their observations, are consistent (in frequency) with two of the three identified by \citet{2003ApJ...591.1184F}.

\subsubsection{\textit{TESS} observations}

\begin{figure*}
\centering
\includegraphics[width=18cm]{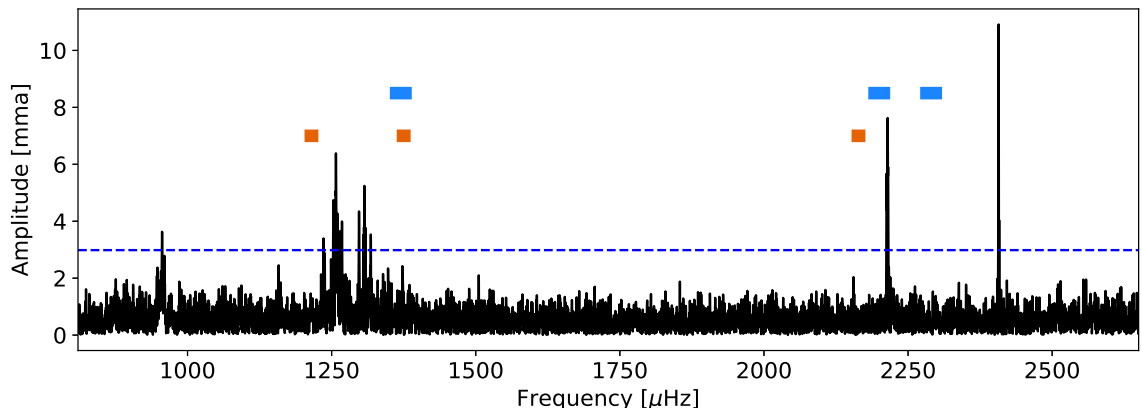}
\caption{Fourier transform of the {\it TESS} light curve of \mct; the blue dashed line indicates the 0.1\% FAP level (S/N = 4.57). Red bars mark the frequencies reported by \citet{2003ApJ...591.1184F}, and blue bars above mark those noted by \citet{2014MNRAS.437.1836K}.}
\label{fig:mctFT}
\end{figure*}

From the {\it TESS} light curve, we see that \mct\ is still pulsating, with periodicities that are in the same range as reported earlier.  However, some differences with the earlier data are evident.  The highest peak in the FT in Fig.~\ref{fig:mctFT} is a single, fully resolved peak at 2407.20~$\mu$Hz, with an amplitude of 11~mma, that was not seen in the earlier work. In this context, a frequency is considered resolved when it is a single peak that can be pre-whitened with a sinusoid. The second-highest peak(s) lie between 2212 and 2216~$\mu$Hz, consistent with what was reported earlier.  There are at least three peaks within this frequency range, but the 20.3 day span of Sector 3 data on this star was not long enough to fully resolve them.  Least-squares fits and pre-whitening in this frequency range revealed at least 3 closely spaced but unresolved modes. 

\begin{figure*}
\centering
\includegraphics[width=18cm]{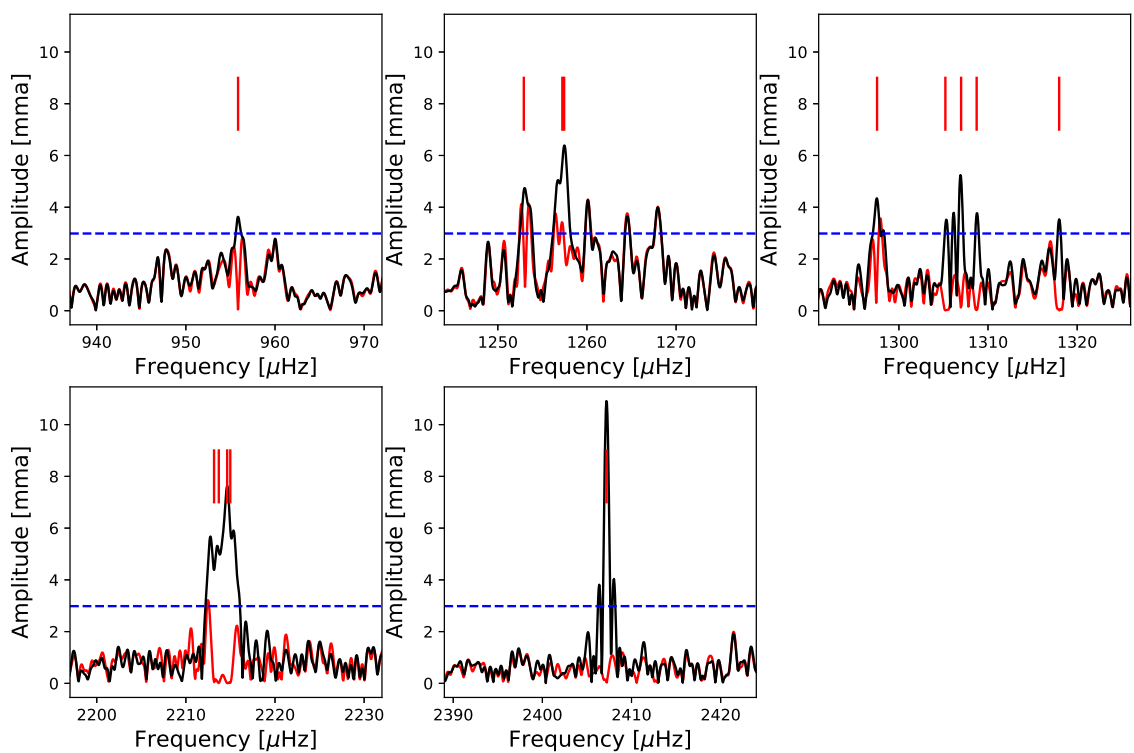}
\caption{Fourier transform of the {\it TESS} light curve of \mct; the blue dashed line indicates the 0.1\% FAP level (S/N = 4.57). Red bars mark the frequencies obtained by a successive least-squares fitting and pre-whitening process. Residuals after pre-whitening are shown as the red curve.}
\label{fig:mctFTz}
\end{figure*}

At lower frequencies, the {\it TESS} data show several groups of significant peaks that fall within roughly Lorentzian envelopes, at 956~$\mu$Hz, 1257~$\mu$Hz and 1306~$\mu$Hz (periods of 1046~s, 796~s, and 766~s).  As with the other cool DAVs in this study, we were unable to completely pre-whiten peaks within these three regions, indicating that the modes may be amplitude/phase unstable over the span of the observations.  In Fig.~\ref{fig:mctFTz}, we expand the view around several peaks, and show the residuals after removing the largest peaks in each range.  This illustrates that the 2407.20~$\mu$Hz mode is a single coherent periodicity, while the modes with frequencies below 1400~$\mu$Hz are complex.  The 2215~$\mu$Hz mode seems under-resolved -- we can find 4 frequencies using least-squares sinusoidal fitting that removes most of the peak, but the frequencies are separated by small amounts that are nearly at the expected resolution for a run of this length.

Summarising, the only resolved, coherent oscillation can be fit with a sinusoid with a frequency of $2407.200 \pm 0.014\,\mu$Hz, with an amplitude of $10.9 \pm 0.6$\,mma. Considering the mode near $2215\,\mu$Hz, there is insufficient time to resolve this into individual peaks, and therefore we can not yet say if this is a stochastically excited mode or if it is a closely spaced multiplet. If it is a closely spaced multiplet, and rotation is what causes the splitting, then it would be one of the most slowly rotation white dwarfs with asteroseismic results as per Hermes et al.~(2017).

There seems to be a `triplet' at 1305.215, 1306.980, and 1308.724\,$\mu$Hz (all about $\pm 0.04$\,$\mu$Hz) though all three are close to the false-alarm limit.  A 1.7\,$\mu$Hz splitting corresponds to a rotation period of 3.3 days if that `triplet' is indeed associated  to an $\ell=1$ mode. The peaks on either side of that `triplet' at 1297.55 and 1317.98 are nearly equally spaced around that 1307\,$\mu$Hz clump. However, the three peaks close to 1307\,$\mu$Hz are not a triplet but a single stochastically excited mode, and the 1298 and 1318\,$\mu$Hz to either side make up the rest of the (more widely spaced) triplet that would be present if the rotation period was 0.57 days. We were unable to get a satisfactory fit using Lorentzians for that wider-triplet case.

As we mentioned above, there are two additional frequency regions with peaks at 956 and 1257\,$\mu$Hz, respectively. Note that they are also not resolved, and all close enough to the significance threshold. 

\begin{table*}
\centering
\caption{Complete list of frequencies, periods, and amplitudes derived from the \mct\ data set.}
\label{tabl:mctfreq}
\begin{tabular}{rrr}
\hline
\hline
\multicolumn{1}{c}{Frequency [$\mu$Hz]} & \multicolumn{1}{c}{Period [s]} & \multicolumn{1}{c}{Amplitude [mma]} \\
\hline
955.854(044) & 1046.185 & 3.68(53) \\
1253.099(045) & 798.021 & 3.63(1.89) \\
1257.085(063) & 795.491 & 5.29(53) \\
1257.462(052) & 795.252 & 6.40(88) \\
1297.472(039) & 770.730 & 4.13(62) \\
1305.221(045) & 766.154 & 3.72(55) \\
1306.979(033) & 765.123 & 5.21(89) \\
1308.725(053) & 764.103 & 3.18(55) \\
1317.983(046) & 758.735 & 3.52(1.90) \\
2213.161(046) & 451.842 & 5.58(63) \\
2213.695(052) & 451.733 & 8.13(54) \\
2214.635(062) & 451.542 & 11.36(54) \\
2214.985(055) & 451.470 & 8.35(53) \\
2407.200(015) & 415.420 & 10.90(53) \\
\hline
\end{tabular}
\end{table*}


\subsection{\lnt}
\begin{figure*}
\centering
\includegraphics[width=16cm]{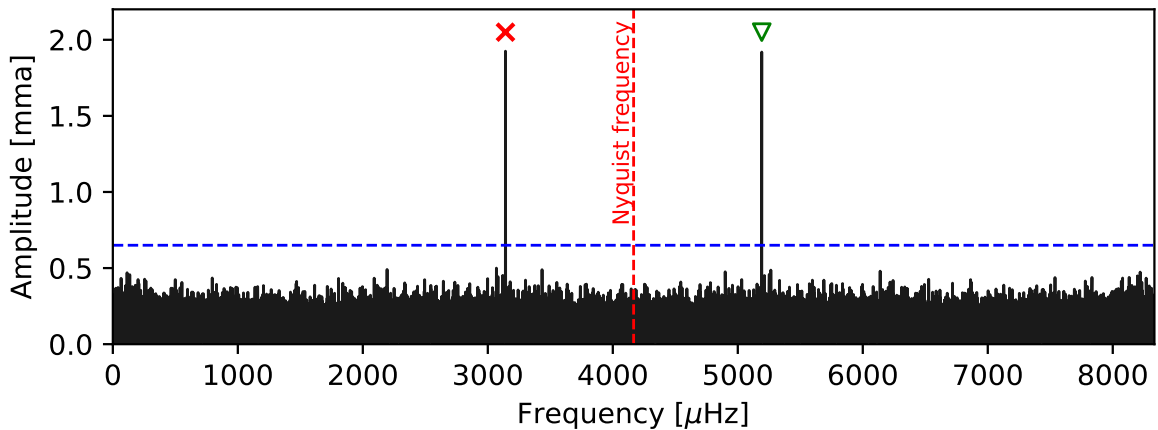}
\caption{Fourier transform of the {\it TESS} light curve of \lnt\ calculated out to twice the Nyquist frequency. The single significant observed signal is reflected across the Nyquist frequency (red dashed line), with the incorrect sub-Nyquist alias marked with a red $\times$ and the signal at the intrinsic frequency marked with a green triangle. The blue dashed line indicates the 0.1\% FAP level.}
\label{fig:l19-2_per}
\end{figure*}

\subsubsection{Ground-based observations}

\lnt\ (TIC\,262872628, $G=13.44$\,mag, $\alpha_{2000}=14^{\mathrm h}33^{\mathrm m}08^{\mathrm s}$, $\delta_{2000}=-81^{\mathrm d}20^{\mathrm m}14^{\mathrm s}$) was discovered by \citet{1977ApJ...215L..75H} to be a relatively simple ZZ~Ceti with a pulsation spectrum dominated by two short-period modes with periods of 193\,s and 114\,s.  This star was the target of a Whole Earth Telescope \citep{1990ApJ...361..309N} campaign in 1995, which revealed \lnt\ to pulsate in at least 10 detectable eigenmodes \citep[preliminary analyses are presented in][]{1998BaltA...7..159S,2005ApJ...635.1239Y,2015ASPC..493..199S}.  Continued monitoring from Mt John University Observatory was used to measure the secular rate of period change from the cooling (and proper motion) of this white dwarf of $dP/dt = (4.0\pm0.6)\time10^{-15}$\,s\,s$^{-1}$ \citep{2015ASPC..493..199S}, the third such measurement for a DAV.  This value is slightly higher than predicted by asteroseismic models, causing \citet{2016JCAP...07..036C} to argue that the cooling could be accelerated by the emission of axions.
\citet{2001ApJ...552..326B} also presented asteroseismological constraints on the structure of \lnt.

\subsubsection{\textit{TESS} observations}

The periodogram of the \textit{TESS} Sector 12 data on \lnt\ reveals a single significant signal at $3141.461\pm0.013\,\mu$Hz with an observed amplitude of $1.92\pm0.11$\,mma.  We recognise this as the alias of the dominant mode observed in previous studies of this star, reflected across the Nyquist frequency.  We identify the intrinsic signal from this mode in the periodogram computed out to twice the Nyquist frequency, as displayed in Figure~\ref{fig:l19-2_per}. Refining the fit of a sinusoid for this signal to the light curve, we obtain a frequency measurement of $5191.853\pm0.013\mu$Hz ($192.6095\pm0.0005$\,s period).  The 2-minute exposures smooth the light curve, reducing the observed amplitude of this signal to 47\% of its intrinsic amplitude.  The amplitude measured in shorter exposures would be close to 4\,mma, which is comparable to values measured from high speed photometry in the literature, allowing for the difference in observational filter.  The second highest amplitude signal previously observed from \lnt\ has a period of 114\,s; this is very close to the \textit{TESS} exposure time, suppressing this signal to well below the periodogram noise floor at 5\% its intrinsic amplitude.

\begin{table*}
\centering
\caption{Frequency derived from the \lnt\ data set.}
\label{tabl:lfreq}
\begin{tabular}{lrr}
\hline
\hline
\multicolumn{1}{c}{Frequency [$\mu$Hz]} & \multicolumn{1}{c}{Period [s]} & \multicolumn{1}{c}{Ampl. [mma]} \\
\hline
5191.853(013) & 192.610 & 1.92(11) \\
\hline
\end{tabular}
\end{table*}


\subsection{\he}

\subsubsection{Ground-based observations}
\he\, (TIC 382303117, $G=15.95$\,mag, $\alpha_{2000}=05^{\mathrm h}33^{\mathrm m}07^{\mathrm s}$, $\delta_{2000}=-56^{\mathrm d}03^{\mathrm m}53^{\mathrm s}$) was discovered in a 2-hour duration high S/N photometric observation \citep{2003ApJ...591.1184F}. \he\ showed two primary periods of oscillation near 688.8 and 586.4\,s (1451.8 and 1705.3~$\mu$Hz), with amplitudes near 8\,mma. 

\citet{2009MNRAS.396.1709C} found this star to be near the red edge of the ZZ Ceti instability strip at $T_{\rm eff}$ = 11\,560\,K, thus it is expected to have complex and long periodic oscillations generally exceeding 600\,s. They identified additional periodicities in the \citet{2003ApJ...591.1184F} data, ranging from 1100--1950\,$\mu$Hz. The most recent observations of HE 0532-5605 came from \citet{2014MNRAS.437.1836K}, who used the Berkeley Visual Imaging Tube detector to observe \he\, for roughly 3000~s with a 0.1~s integration time. They detected periods of 438, 707, and 1380\,s (720, 1410, and 2280~$\mu$Hz respectively) and indicated that they believed that the periodicity around 1410~$\mu$Hz may be unstable and related to the peak originally detected by \citet{2003ApJ...591.1184F}.

\subsubsection{\textit{TESS} observations}
\begin{figure*}
\centering
\includegraphics[width=18cm]{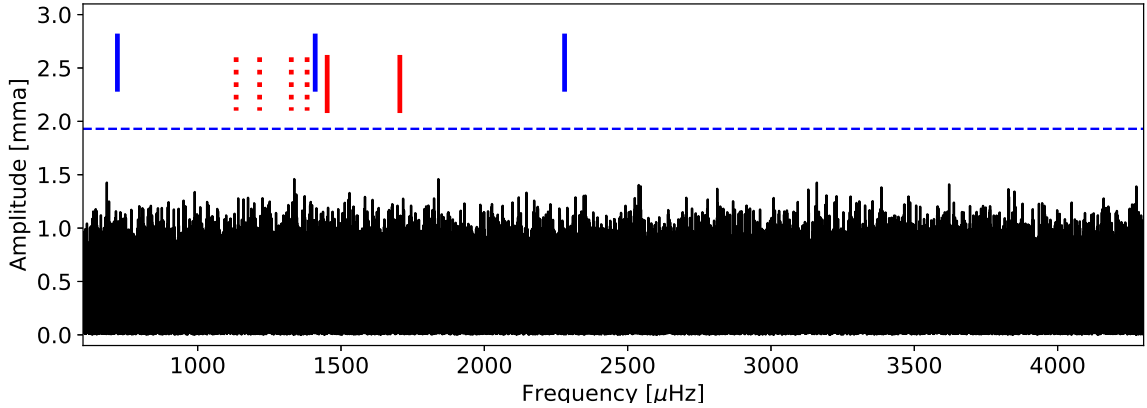}
\caption{Fourier transform of Sectors 1--13 of the {\it TESS} light curve of \he; the blue dashed line indicates the 0.1\% FAP level (S/N = 5.24). Vertical red lines mark the frequencies from \citet{2003ApJ...591.1184F}, dashed lines are additional frequencies from that data reported by \citet{2009MNRAS.396.1709C}, and blue vertical lines (offset upwards) mark those from \citet{2014MNRAS.437.1836K}.}
\label{fig:heFT}
\end{figure*}

\he\, is located near the southern ecliptic pole and as a result can be observed by {\it TESS} for nearly a year. Here, we report on data from all of the 13 sectors available. Each sector was reduced independently, as described earlier, after which the sectors were combined into a single light curve.  The Fourier transform of the entire data set is shown in Fig.~\ref{fig:heFT} -- the 0.1\% FAP level (S/N = 5.24) lies at 1.93 ppt, which is remarkable given the faintness of the star and the limited aperture of the {\it TESS} camera. There are no significant periodicities detected in \he\, in the complete data set.

The amplitudes of the modes reported from ground-based studies ranged from 2--8\,mma, and were seen in much shorter data sets than the full {\it TESS} light curve.  Given that the reported periods are in excess of 600\,s, and that this is a relatively cool DAV, it is likely that the modes observed from the ground have short lifetimes resulting from changes in phase and amplitude.  As we saw in other targets in this study, and described by \citet{2015ApJ...809...14B}, this lack of coherence can reduce the signal amplitude in the Fourier transform if the duration of the observations is much longer than the damping time.  Scaling down the expected amplitudes to account for the redder bandwidth of the {\it TESS} data can also help explain the failure of these modes to appear in Fig.~\ref{fig:heFT}.  

\begin{figure*}
\centering
\includegraphics[width=12cm]{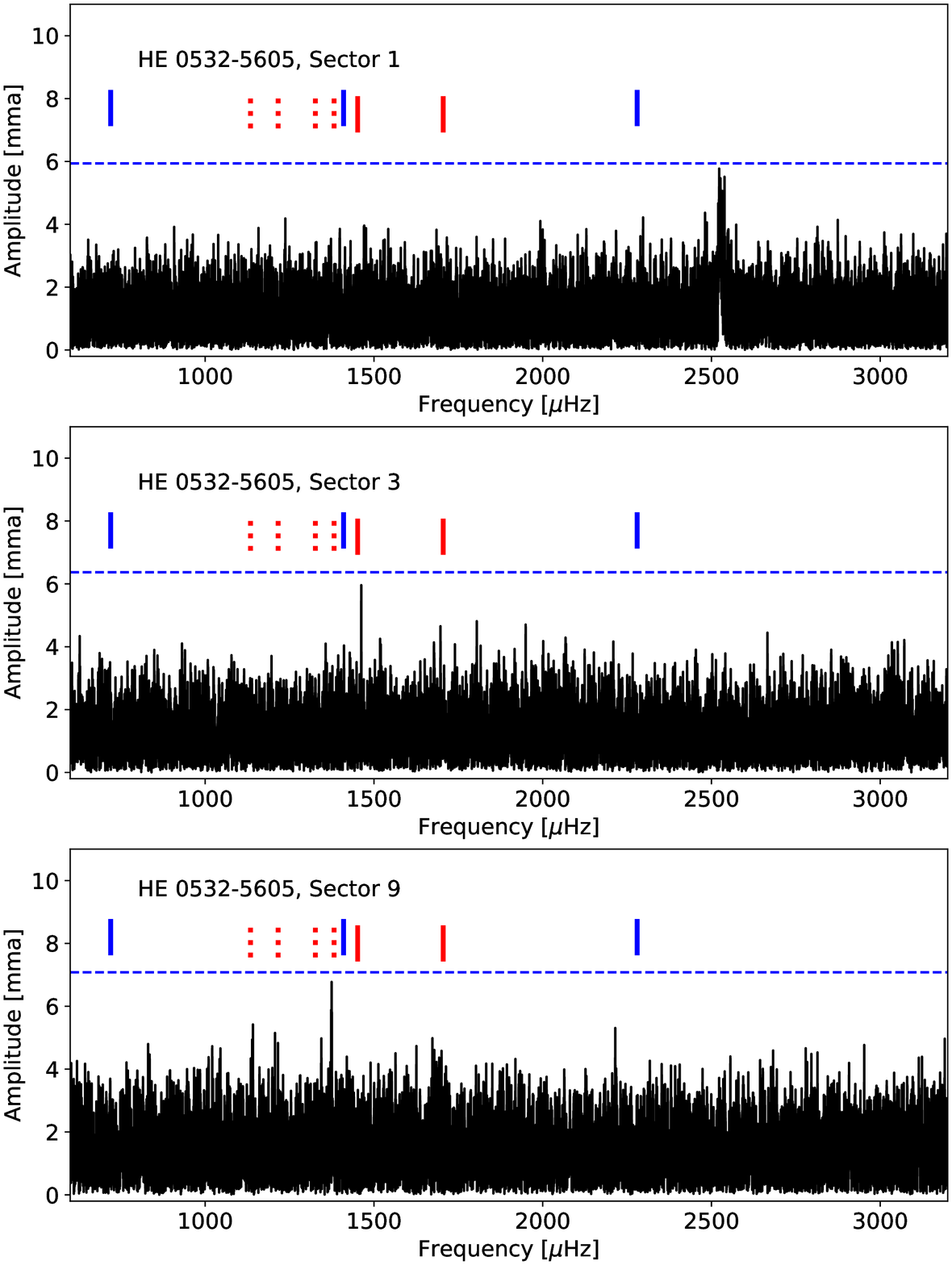}
\caption{Fourier transform of three sectors (labelled) of \he\ data; the blue dashed line indicates the 0.1\% FAP level for each sector. Vertical lines are as in Fig.~\ref{fig:heFT}. Other sectors showed no peaks approaching the 0.1\% FAP limit.}
\label{fig:heFT139}
\end{figure*}

To look for shorter-lifetime modes, we examined each sector of the {\it TESS} data individually to see if significant periodicities were present.  However, since each sector is 1/13 as long as the total run, the noise in the Fourier transform in the individual sectors is nearly three times higher, putting our 0.1\% FAP threshold at about 6\,mma. No significant periodicities were detected, however, in Sectors 1, 3, and 9, some peaks came close to reaching the 0.1\% FAP threshold.  We show Fourier transforms of these three sectors in Fig.~\ref{fig:heFT139}. In Sectors 3 and 9, the highest peaks were close to frequencies reported by \citet{2003ApJ...591.1184F} and \citet{2009MNRAS.396.1709C}.

\begin{figure*}
\centering
\includegraphics[width=18cm]{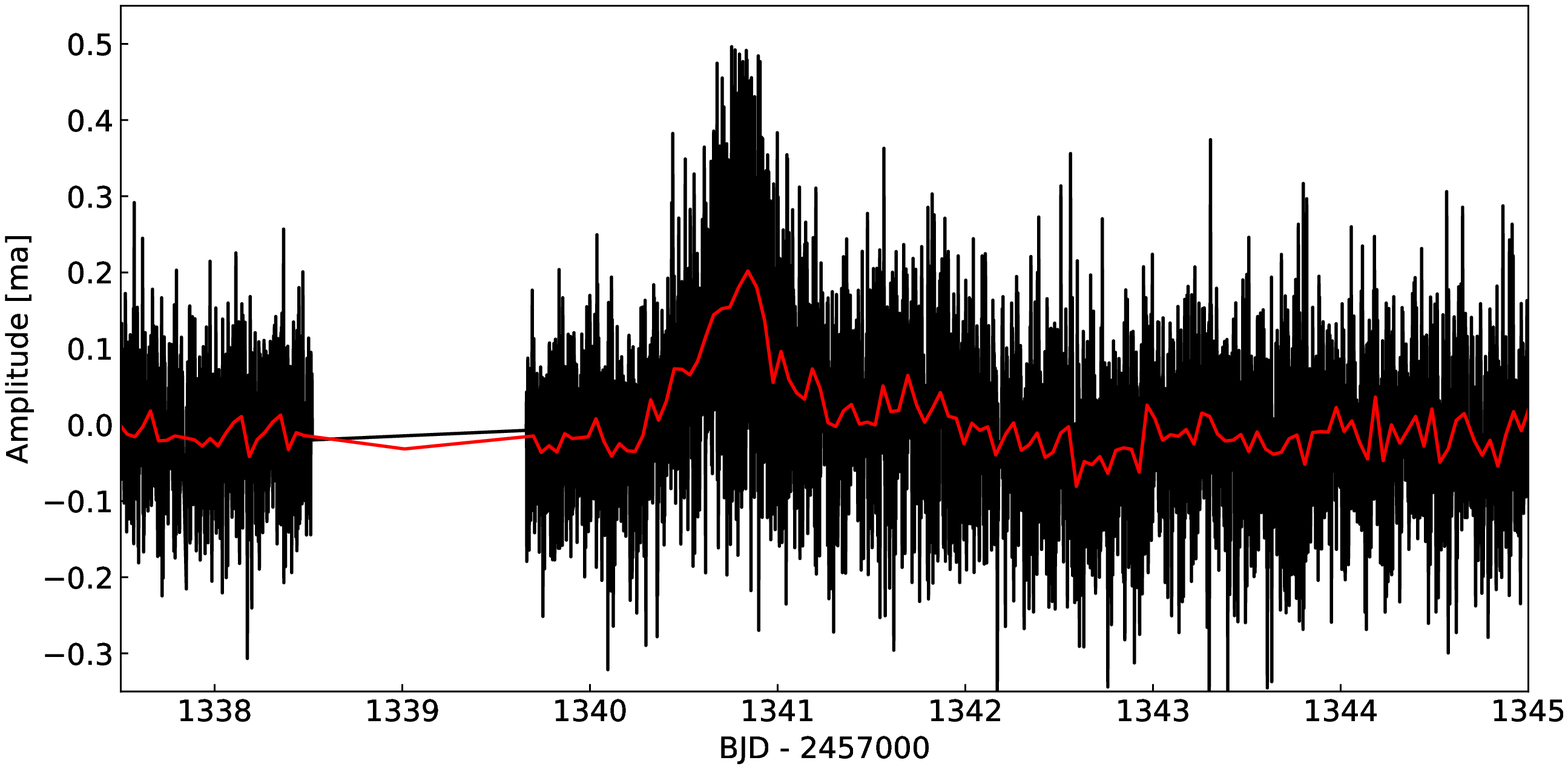}
\caption{Light curve of a portion of the Sector 1 observations of \he. The red line is binned light curve, with a bin size of 30 points, to show the amplitude and duration of an apparent brightening of 20\% for 24 hours.}
\label{fig:hes1lc}
\end{figure*}

The peak near 2500~$\mu$Hz in Sector 1 is broad and reminiscent of those seen in the shorter runs on cool DAV stars here \citep[and in][]{2017ApJS..232...23H} and in {\it K2} observations of the  DAV PG~1149+057 \citep{2015ApJ...810L...5H}.  In the latter case, the star in question is a ``burster'', which shows semiregular increases in brightness coincident with large amplitude pulsations.  Close examination of the Sector 1 light curve of \he\, suggests that such a burst may have occurred at about BJD~2\,458\,340.  In Fig.~\ref{fig:hes1lc} we expand the light curve around this interval.  For this figure, we did not flatten the light curve, as the flattening process reduced the amplitude of the apparent burst.  The possible burst reaches an amplitude of nearly 20\%, and the duration is approximately 24 hours.

\begin{figure*}
\centering
\includegraphics[width=18cm]{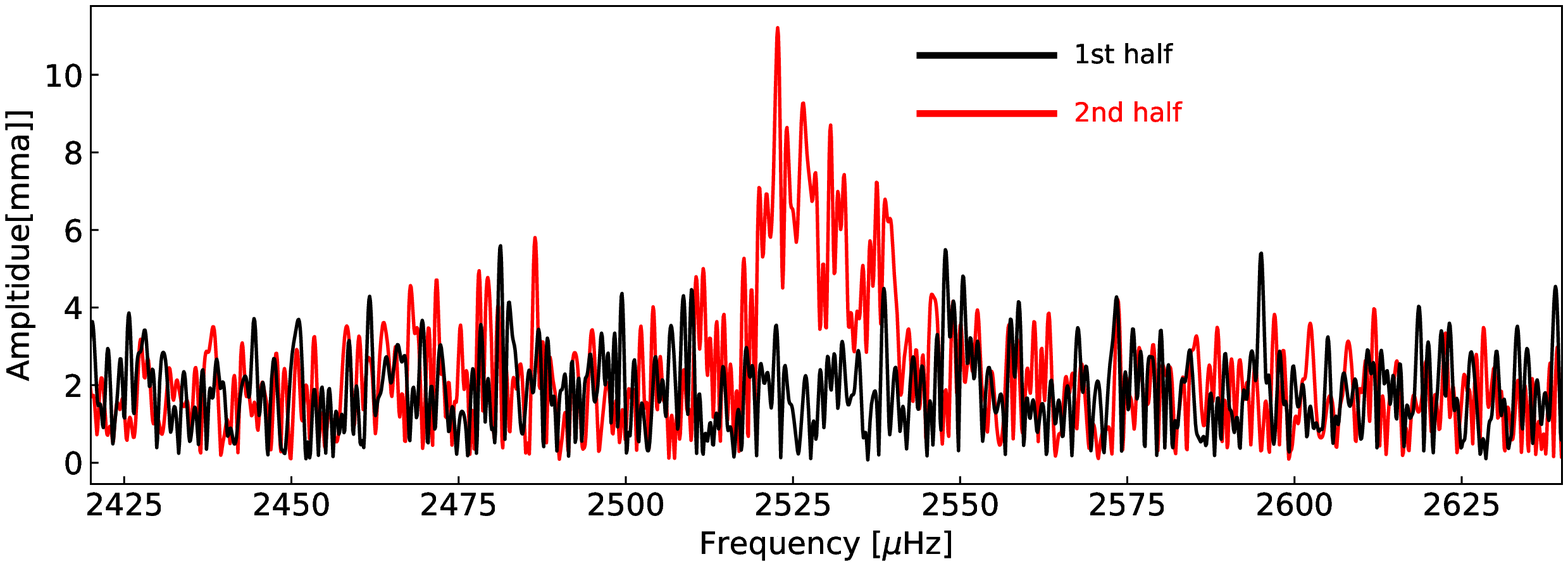}
\caption{Fourier transform of Sector 1 data on \he. The black curve is for the first half of Sector 1, and the red curve is for the second half, which contains the candidate burst}
\label{fig:hes1ft}
\end{figure*}

The feature at 2530~$\mu$Hz in the top panel of Fig.~\ref{fig:heFT139} appears to be associated (in time) with the feature at 1341~d (BJD - 2\,457\,000) in Fig.~\ref{fig:hes1lc}.  Figure~\ref{fig:hes1ft} shows the Fourier transform of the first and second halves of the Sector 1 data on \he.  The burst appears soon after the start of the second half of the observation.  The feature at 2530~$\mu$Hz is only present in the second half of the sector as well. 

Since this star only showed mildly coherent pulsations during a portion of Sector 1, determining the frequency or frequencies present is a very uncertain task.  The FT rose above the significance threshold between approximately 2520 and 2550\,$\mu$Hz.  We fit a Lorentzian envelope to the strongest part of this range and so estimate the frequency for this mode at approximately 2527\,$\mu$Hz.

It appears, therefore, that \he\, may be a bursting, pulsating DA white dwarf. The duration and amplitude of the burst, and the increased pulsation amplitude during the burst, has been seen in other stars.  However, most of the bursting white dwarfs recur on an irregular basis, but roughly every 8 days or so.  No other clear bursts have been seen in the {\it TESS} data over 13 sectors, though it is possible that the nearly-significant features in Sectors 3 and 9 may reflect smaller bursts that aren't apparent in the relatively low S/N data from this faint star. Also, the periodicity present during the burst is at a relatively short period (395~s, or 2530~$\mu$Hz). \he\,  merits monitoring with larger instruments to confirm its possible membership in the rare class of bursting, pulsating DA white dwarfs.


\subsection{\hs}

\subsubsection{Ground-based observations}

\hs\ (TIC\,455094688, $G=15.40$\,mag, $\alpha_{2000}=05^{\mathrm h}10^{\mathrm m}14^{\mathrm s}$, $\delta_{2000}=+04^{\mathrm d}38^{\mathrm m}55^{\mathrm s}$) was discovered to be a member of the ZZ~Ceti class by \citet{1998A&A...330..277J}. Their observations showed that the star's light variations can be described with three or four independent modes at 278.4, 355.1, 445.2, and 558.7\,s, and a large number of combination terms, with 18 frequencies altogether. Further observations revealed that at some frequencies there are actually rotationally split triplets. \citet{2002MNRAS.335..399H} detected 10 independent frequencies consisting of three triplets and a singlet frequency, and also a large number of combination frequencies, while \citet{2002A&A...388..219K} detected two doublets and three additional modes besides the combination terms. \citet{2009MNRAS.396.1709C} listed four modes for their asteroseismic fit, while the latest time-series photometric observations on \hs\ were presented by \citet{2013MNRAS.429.1585F}. They derived 18 independent pulsation frequencies consisting of five triplets, a doublet, and a singlet frequency besides the combination terms. 

\subsubsection{\textit{TESS} observations}

\begin{table*}
\centering
\caption{Six-frequency solution of the \hs\ data set. The errors are formal uncertainties calculated by \textsc{famias}. In the case of the rotationally split frequencies, we mark the azimuthal order values, based on the results of the subsequent observations.}
\label{tabl:hsfreq}
\begin{tabular}{lrrr}
\hline
\hline
 & \multicolumn{1}{c}{Frequency [$\mu$Hz]} & \multicolumn{1}{c}{Period [s]} & \multicolumn{1}{c}{Amplitude [mma]} \\
\hline
$f_1$ & 1680.072(32)	& 595.213	& 5.42(71) \\
$f_2^{-1}$ & 2242.401(13)	& 445.950	& 13.08(71) \\
$f_2^{+1}$ & 2248.078(28)	& 444.824	& 6.07(71) \\
$f_3^{-1}$ & 2810.935(12)	& 355.754	& 14.66(72) \\
($f_3^{-0}$ & 2814.347(52)	& 355.322	& 3.29(72) \\
$f_3^{+1}$ & 2817.615(18)	& 354.910	& 9.33(72) \\
\hline
\end{tabular}
\end{table*}

\begin{figure*}
\centering
\includegraphics[width=18cm]{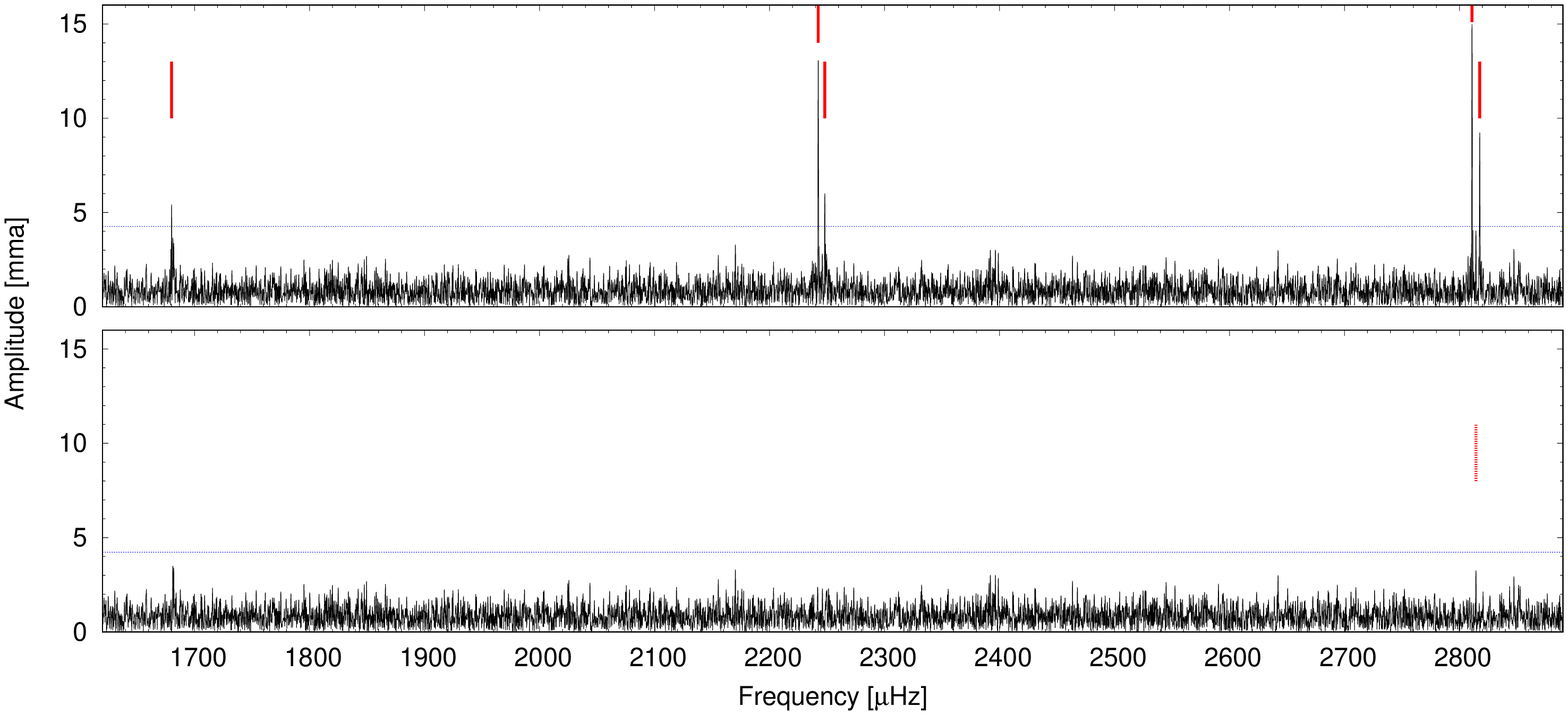}
\caption{Fourier transform of the light curve of \hs\ before (upper panel) and after (lower panel) pre-whitening with the five significant frequencies. Red dashed lines mark the frequencies listed in Table~\ref{tabl:hsfreq}, while the blue dashed lines represent the $0.1\%$ FAP levels (S/N=4.73).}
\label{fig:hsfreq}
\end{figure*}

In contrary to the rich frequency spectrum observed e.g. by \citet{2013MNRAS.429.1585F}, we only detected five peaks in the \textit{TESS} data set of \hs\ above the significance level. Table~\ref{tabl:hsfreq} lists the frequencies, periods and amplitudes of these peaks. The $1680.072\,\mu$Hz frequency represents a new mode not observed before, while the other four peaks are side components of formerly detected triplets. No combination frequencies could be derived by the \textit{TESS} data, however, we have to take into account the amplitude suppression factor of the 2-minute exposures, especially in the case of peaks expected to emerge with low amplitudes. Note that we listed an additional peak in Table~\ref{tabl:hsfreq} in parentheses. This represents the central component of the $\sim 2810\,\mu$Hz triplet, however, this peak is under our significance threshold (S/N=3.7).

The $f_2$ and $f_3$ frequencies are found to be stable during the time-span of the \textit{TESS} observations, as these frequency domains can be pre-whitened with the doublet frequencies alone, and no further closely spaced frequencies emerged as in e.g. \ec. Figure~\ref{fig:hsfreq} shows the original and the pre-whitened Fourier transforms, respectively.

Considering the influence of the companion HS~0507+0434A (TIC~455094685) on the frequency determination of \hs, we found that \textit{TESS} cannot separate individual signals from these two objects since the separation between the two stars ($\sim15$\,arcsec) is smaller than the plate scale of the \textit{TESS} detector (21\,arcsec\,px$^{-1}$). However, investigating the amplitudes of the signals originate from the two sources, we found that at the position of \hs\ the measured amplitudes are a factor of $4.02 \pm 0.04$ (error of the mean) larger than at the position of HS 0507+0434A, so they must come from component B. Another argument that the frequencies detected belong to \hs, is that there are ground-based studies which clearly resolve both components, including the paper of \citet{2002MNRAS.335..399H}, where the authors studied the variability of HS~0507+0434B, using HS~0507+0434A as well as other stars for comparison.  


\subsection{ZZ~Ceti stars not observed to vary in the \textit{TESS} data}

\textit{TESS} targeted additional previously known, bright ZZ Ceti variables in the southern hemisphere that did not exhibit significant pulsation signals in the \textit{TESS} data, namely MCT\,2148$-$2911, HE\,0031$-$5525, EC\,00497$-$4723, MCT\,0016$-$2553, WD\,0108$-$001, HS\,0235+0655, KUV\,03442+0719, WD\,J0925+0509, HS\,1013+0321, and EC\,11266$-$2217. For completeness, we report on these ten stars in Appendix~\ref{app:nov}, comparing the upper-limit detection thresholds of the \textit{TESS} data to pulsation amplitudes reported in the literature.


\section{Summary and conclusions}
\label{sect:disc}

We presented the frequency analysis of 18 formerly known ZZ~Ceti stars observed by the \textit{TESS} space telescope during the survey observation of the southern ecliptic hemisphere (sectors 1--13). We compared our results to those from previous ground-based observing campaigns.
Eight out of 18 of our targets show at least marginal evidence of pulsations that we were able to characterise, with at least 40 statistically significant (FAP < 0.1\%) pulsation detections in total. From $\sim$month-long \textit{TESS} light curves with very high duty cycles, we have measured pulsation frequencies to $\lesssim 0.1\,\mu$Hz precision.  These are eigenfrequencies of these white dwarf stars and they can be compared to stellar models to infer the interior conditions of these compact pulsators.  A follow-up paper from TASC WG\#8 is planned (Romero et al.\ in prep.) that will use different sets of models to interpret the precision measurements reported here.

\begin{figure*}
\centering
\includegraphics[width=13cm]{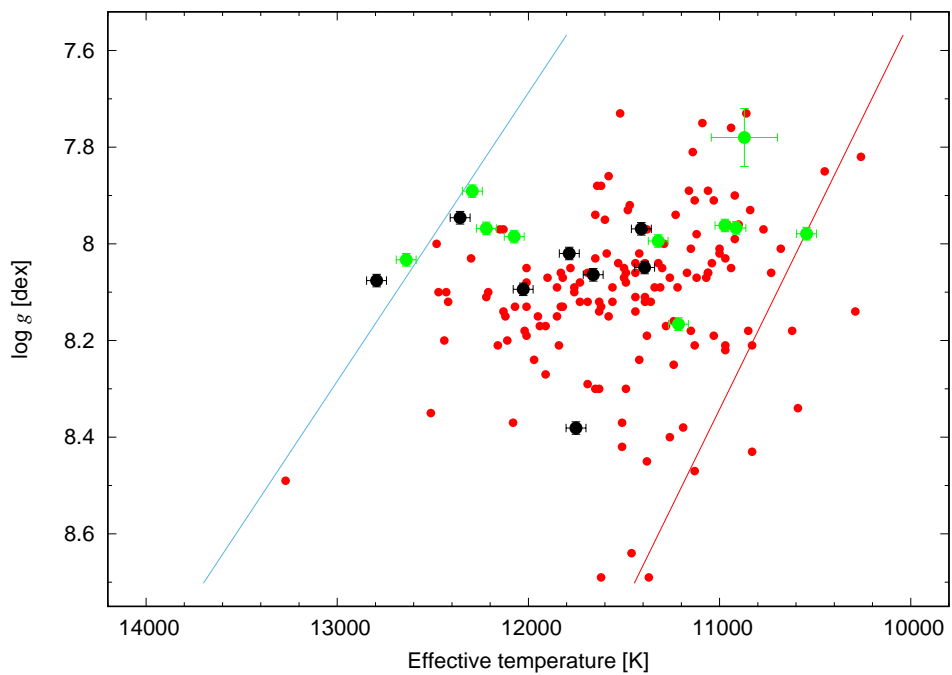}
\caption{Known ZZ~Ceti stars (red filled dots, atmospheric parameters from \citealt{2016IBVS.6184....1B}), and the DAVs observed by \textit{TESS} during the the survey measurements of the southern ecliptic hemisphere (sectors 1--13) in the $T_{\mathrm{eff}} - \mathrm{log}g$ diagram. Black and green dots mark the stars found to be variable and not observed to vary by the \textit{TESS} measurements, respectively. Their atmospheric parameters were derived by \citet{2017PhDT........20F}, except in the case of KUV~03442+0719, where we used the values presented in \citet{2016IBVS.6184....1B}. Blue and red lines denote the hot and cool boundaries of the instability strip, according to \citet{2015ApJ...809..148T}.}
\label{fig:hrd}
\end{figure*}

Figure~\ref{fig:hrd} shows the ZZ~Ceti instability strip with known ZZ~Ceti stars (atmospheric parameters are from \citealt{2016IBVS.6184....1B}), together with
all the 18 objects studied in this paper. The corresponding atmospheric parameters from the literature are listed in Table~\ref{tabl:hrd2} for the 18 targets \citep{2017PhDT........20F}. The $T_{\mathrm{eff}}$ and $\log g$ values were determined by the use of the ML2$/\alpha=0.8$ version of the mixing-length theory, and corrected according to the findings of \citet{2013A&A...559A.104T} based on radiation-hydrodynamics three-dimensional simulations of convective DA stellar atmospheres.

We demonstrated the difference between the \textit{TESS} amplitudes and those detected in the ground-based measurements through the case of \ross. We have to take several different effects into account. First, \textit{TESS} observes in a redder bandpass than typical ground-based observations, which results in lower pulsation amplitudes, see Eq.~\ref{eg:ross}. Second, the relatively large pixel size of \textit{TESS} often causes contamination, and hence signal-to-noise suppression. Third, the 120\,s exposure times of the \textit{TESS} observations smear out the short-period signals of the ZZ~Ceti stars. And finally, phase and/or amplitude variations on timescales shorter than the month-long \textit{TESS} light curves reduce the pulsation amplitudes by distributing power across multiple frequency bins. These effects explain why we did not detect pulsations in 10 low-amplitude pulsators, and why there are missing frequencies compared to the ground-based results in some of the \textit{TESS} variables.

\begin{figure}
\centering
\includegraphics[width=8cm]{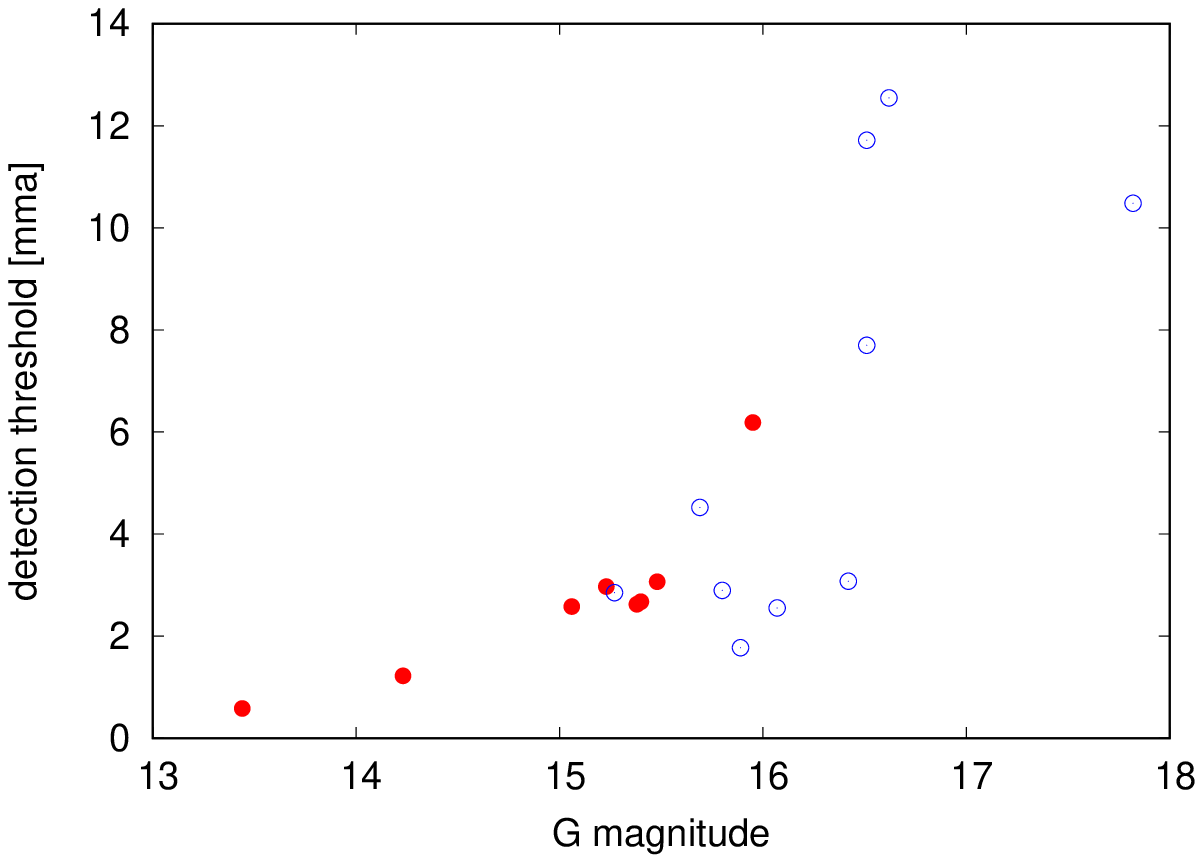}
\caption{Pulsation-detection performance of \textit{TESS}. Stars confirmed (red dots) or not observed to vary (blue circles) by the \textit{TESS} observations in the $G$ magnitude -- detection threshold diagram, where the detection threshold is defined as $0.1\% \mathrm{FAP} \times \sqrt{\mathrm{CROWDSAP}} \times \sqrt{\mathrm{number\,of\,sectors}}$.}
\label{fig:fap}
\end{figure}

The detection threshold corresponding to an uncontaminated target in a single sector is defined as $0.1\% \mathrm{FAP} \times \sqrt{\mathrm{CROWDSAP}} \times \sqrt{\mathrm{number\,of\,sectors}}$ and is shown in Fig.~\ref{fig:fap}. The plot demonstrates the pulsation-detection performance of \textit{TESS}. This threshold obviously depends on the magnitude of the ZZ~Ceti star. 
The eight stars with confirmed pulsations have $G<15.5$\,mag (except for \he). The brightest of the stars with unconfirmed pulsations is $G=15.3$\,mag and most are fainter than $G=15.8$\,mag. There is, however, significant scatter in the calculated threshold at a given brightness, especially at targets fainter than 15.5\,mag, possibly due to instrumental effects.

In spite of the difficulties, we were able to detect new frequencies for five stars (\ec, \bpm, \bpmb, \mct, \hs). We found that \he\ may be a new outbursting ZZ~Ceti star. We also found possible amplitude/phase variations during the \textit{TESS} observations, which resulted in the emergence of groups of peaks in the data set of \ec, \bpmb, and \mct. We fitted Lorentzians to these frequency groups to approximate pulsation frequency values. Such behaviour in these stars was not identified from the ground before.

Another thing we can learn from the \textit{TESS} measurements is that we have to be cautious with the interpretation of the frequencies considering the Nyquist-limit of the 120\,s exposures ($\approx4167\,\mu$Hz). In the case of the peaks close to the Nyquist frequency, we have to take into account the possibility that these are intrinsically above the Nyquist-limit. The barycentric correction applied to the observation timings makes the data deviate from strict equidistancy that, in turn, lifts Nyquist-degeneracy, that is, the pseudo-Nyquist alias will have lower amplitude in the Fourier spectrum than the real frequency peak \citep{2013MNRAS.430.2986M}. However, in the one-month-long \textit{TESS} observations, this effect is far too weak for safe discrimination. We found that the amplitudes across $n \times$Nyquist behave differently depending on a star. In some stars, the amplitudes are the highest in the sub-Nyquist region, getting lower with increasing frequency, however, sometimes we see that the amplitudes of the signals in question are changing significantly up and down, while being higher in the super-Nyquist region across many times the Nyquist frequency. Some of this behaviour could be cause by intrinsic variations of the pulsation signals.

\begin{table}
\centering
\caption{Atmospheric parameters of the ZZ~Ceti stars observed by \textit{TESS} during the the survey measurements of the southern ecliptic hemisphere (sectors 1--13). The $T_{\mathrm{eff}}$ and $\mathrm{log}g$ values are from \citet{2017PhDT........20F}, except for KUV~03442+0719, where we used the values listed in \citet{2016IBVS.6184....1B}. The external uncertainties for the observations presented by \citet{2017PhDT........20F} are 52\,K in $T_{\mathrm{eff}}$ and 0.013\,dex in $\mathrm{log}g$, respectively, while 173\,K and 0.06\,dex for KUV~03442+0719, as given by \citet{2011ApJ...743..138G}. 
}
\label{tabl:hrd2}
\begin{tabular}{lcc}
\hline
\hline
Object & $T_{\mathrm{eff}}$ & $\mathrm{log}g$\\
 & [K] & [dex] \\
\hline
Ross 548 & 12\,357 & 7.946 \\
EC 23487-2424 & 11\,409 & 7.969 \\
BPM 31594 & 11\,786 & 8.020 \\
BPM 30551 & 11\,392 & 8.049 \\
MCT 0145-2211 & 11\,661 & 8.064 \\
L 19-2 & 12\,794 & 8.076 \\
HE 0532-5605 & 11\,751 & 8.381 \\
HS 0507+0434B & 12\,027 & 8.094 \\
\\
MCT 2148-2911 & 12\,220 & 7.968 \\
EC 00497-4723 & 11\,321 & 7.994 \\
EC 11266-2217 & 12\,074 & 7.985 \\
MCT 0016-2553 & 10\,972 & 7.962 \\
HS 1013+0321 & 12\,639 & 8.033 \\
HE 0031-5525 & 12\,293 & 7.891 \\
WD J0925+0509 & 11\,215 & 8.166 \\
WD 0108-001 & 10\,545 & 7.979 \\
HS 0235+0655 & 10\,914 & 7.967 \\
KUV 03442+0719 & 10\,870 & 7.78 \\
\hline
\end{tabular}
\end{table}

Despite these difficulties, \textit{TESS} observations of ZZ~Ceti stars proved that measurements of bright DAVs allow us to detect the signs of short-term amplitude/phase variations, and also derive new pulsation modes with high precision thanks to the (almost) continuous, homogeneous and high-quality data. However, the value of 20\,s cadence observations is also obvious. The 2\,min cadence induces a significant decrease of amplitudes, which is largely responsible for the non-detection of pulsations in several targets. Offering 20\,s cadence mode observations for short period compact pulsators in the extended mission could infer a breakthrough in the study of these kinds of variables.  


\begin{acknowledgements}
The authors thank the anonymous referee for the constructive comments and recommendations on the manuscript.
We also thank the comments of Kosmas Gazeas (National and Kapodistrian University of Athens).
This paper includes data collected with the TESS mission, obtained from the MAST data archive at the Space Telescope Science Institute (STScI). Funding for the TESS mission is provided by the NASA Explorer Program. STScI is operated by the Association of Universities for Research in Astronomy, Inc., under NASA contract NAS 5–26555.
Support for this work was provided by NASA through the TESS Guest Investigator program through grant 80NSSC19K0378.
ZsB acknowledges the support provided from the National Research, Development and Innovation Fund of Hungary, financed under the PD$_{17}$ funding scheme, project no. PD-123910. This project has been supported by the Lend\"ulet Program of the Hungarian Academy of Sciences, project No. LP2018-7/2019.
KJB is supported by the National Science Foundation under Award No.\ AST-1903828.
ASB gratefully acknowledges financial support from the Polish National Science Center under project No.\,UMO-2017/26/E/ST9/00703.
GH acknowledges financial support by the Polish NCN grant 2015/18/A/ST9/00578.
RR acknowledges funding by the German Science foundation (DFG) through  grants HE1356/71-1 and IR190/1-1.
WZ acknowledges the support from the Beijing Natural Science Foundation (No. 1194023) and the National Natural Science Foundation of China (NSFC) through the grant 11903005.
\end{acknowledgements}



\bibliographystyle{aa} 
\bibliography{18DAVs} 

\begin{appendix}
\section{\textit{TESS} light curves of stars that are confirmed to vary.}

\begin{figure*}
\centering
\includegraphics[width=20cm]{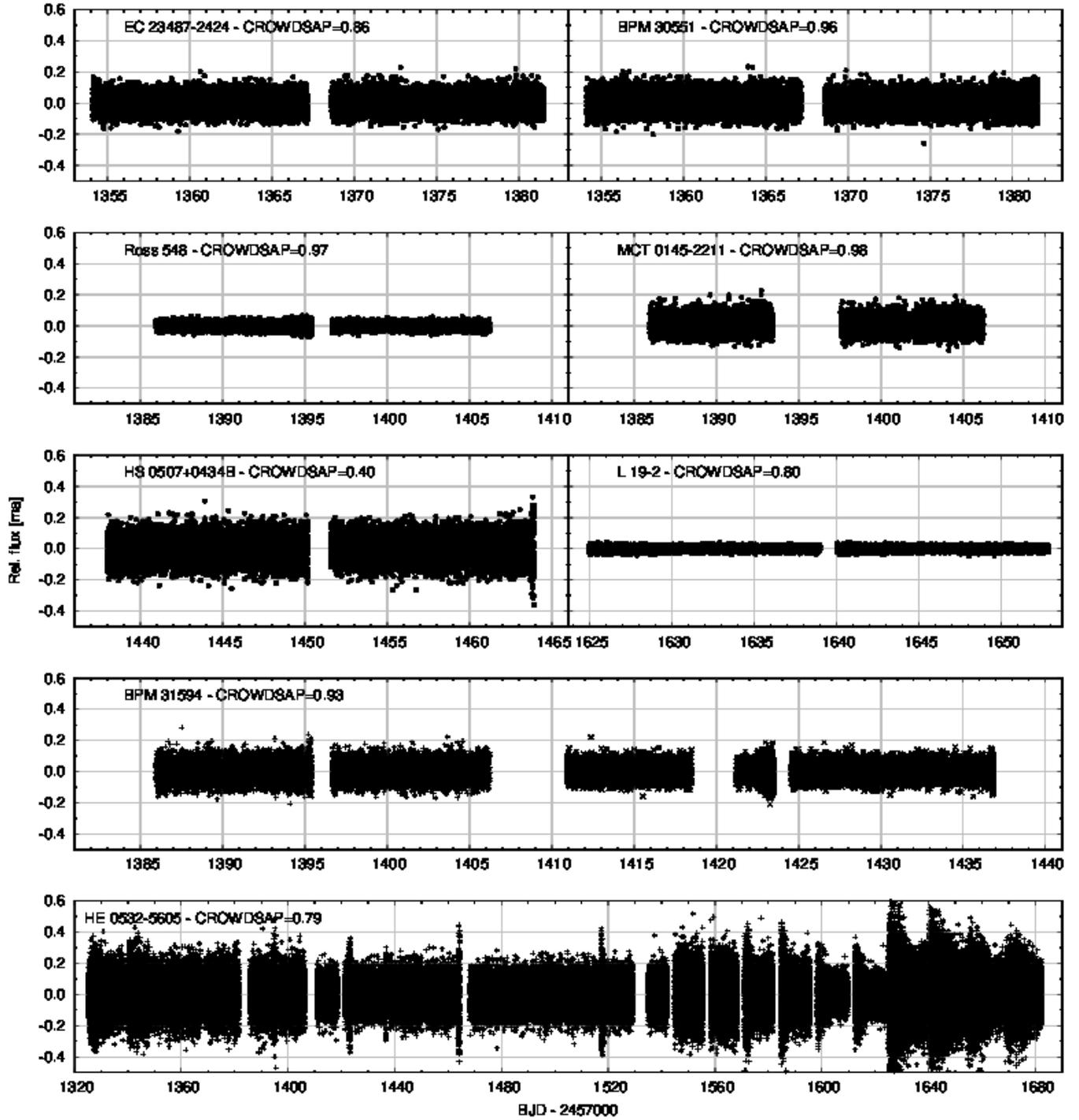}
\caption{Reduced \textit{TESS} light curves of the eight ZZ~Ceti stars found to be variables by the \textit{TESS} data. Note that the light curve of \he\ is on a different scale, and represents 13 sectors of data. We also denoted the values of the CROWDSAP keywords listed in the fits headers, which represent the ratio of the target flux to the total flux in the aperture.}
\label{fig:lcvar}
\end{figure*}

\textit{TESS} light curves of stars that are confirmed to vary, see Fig.~\ref{fig:lcvar}.

\section{ZZ~Ceti stars not observed to vary.}
\label{app:nov}

\begin{figure*}
\centering
\includegraphics[width=20cm]{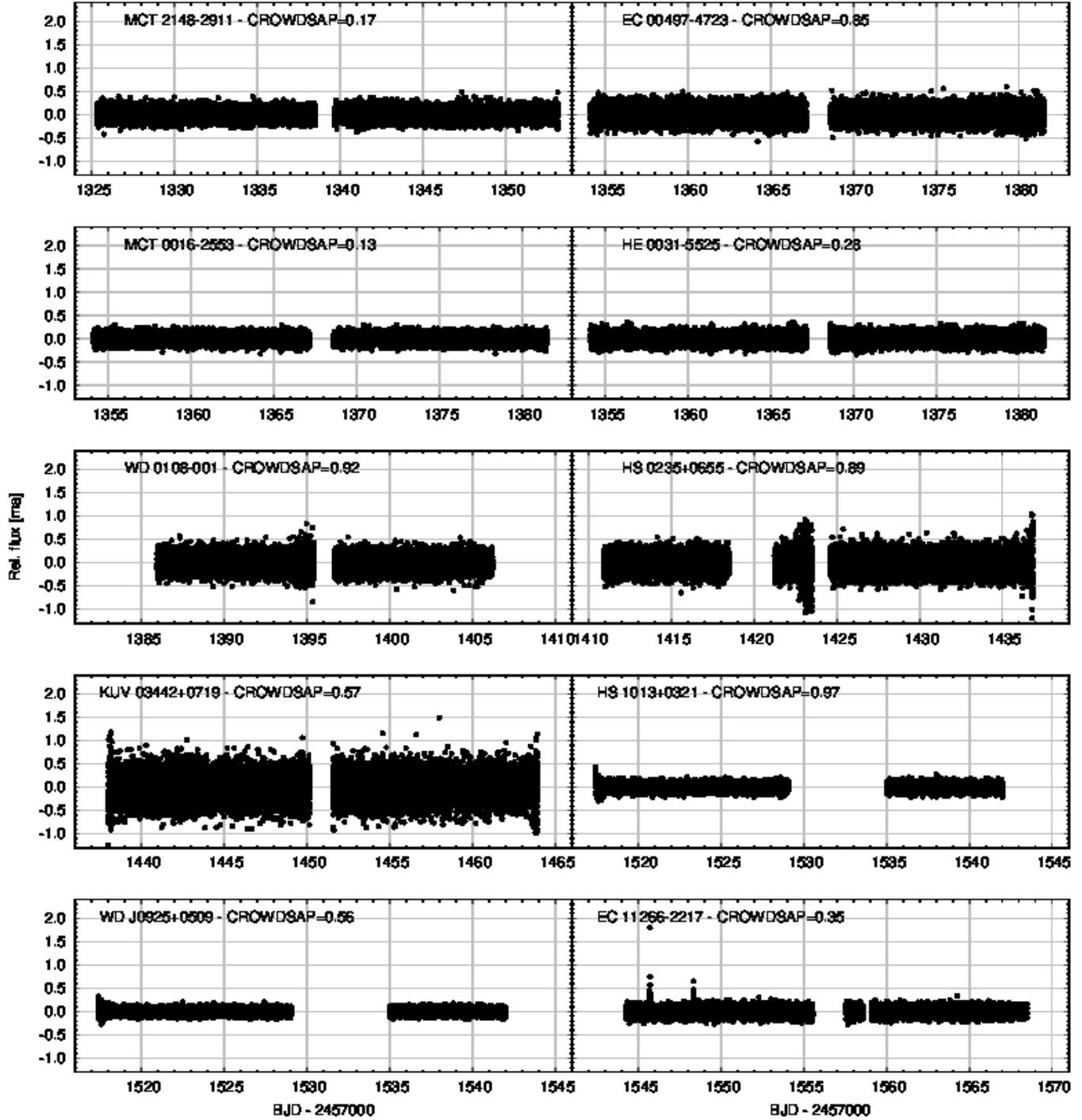}
\caption{Reduced \textit{TESS} light curves of the ZZ~Ceti stars not observed to vary.}
\label{fig:lcnov}
\end{figure*}

\begin{figure*}
\centering
\includegraphics[width=20cm]{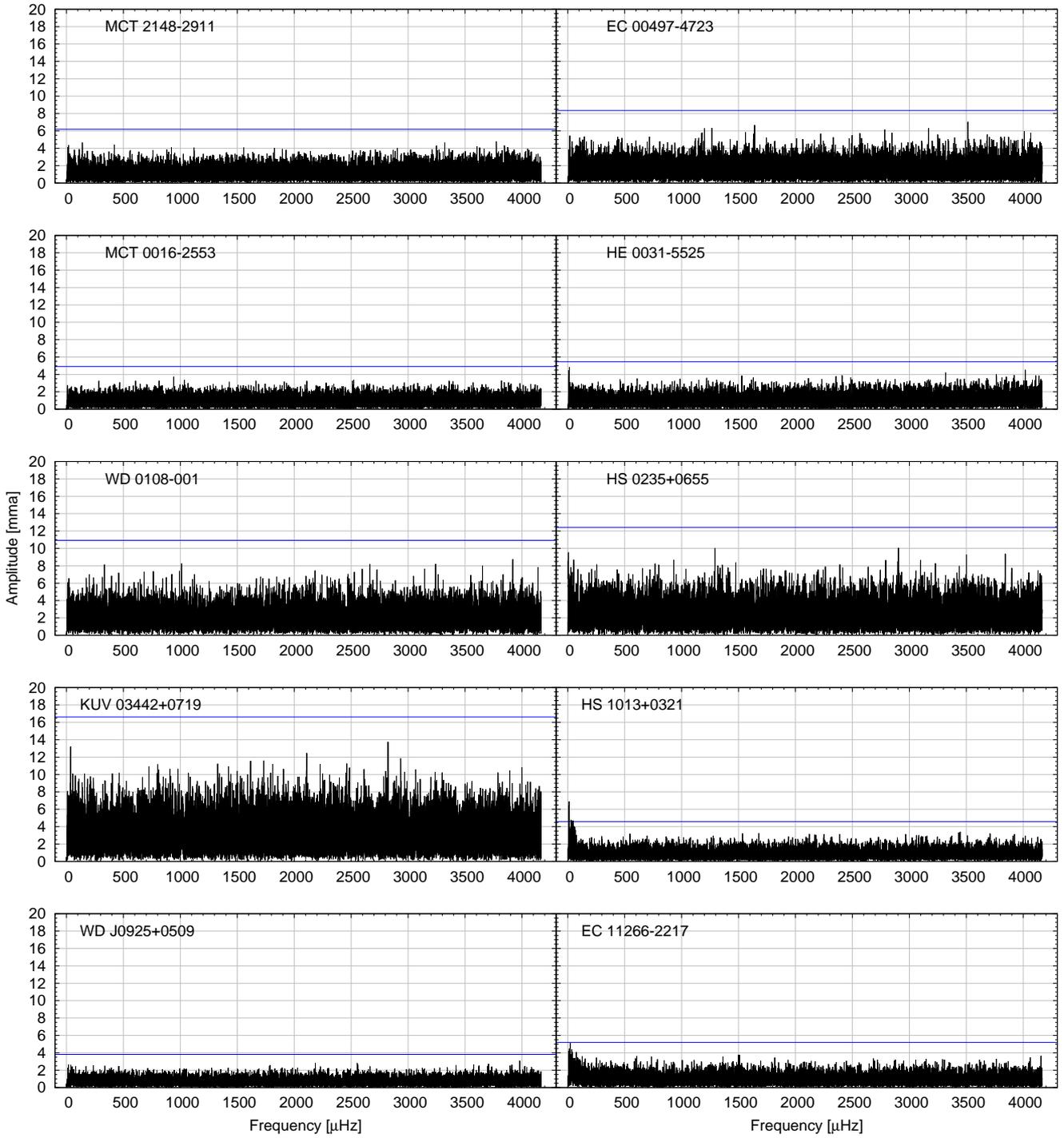}
\caption{Fourier transforms of the ZZ~Ceti stars not observed to vary by the \textit{TESS} data. Blue lines denote the $0.1\%$\,FAP levels.}
\label{fig:spnov}
\end{figure*}

Figure~\ref{fig:lcnov} shows the \textit{TESS} light curves of the stars not observed to vary, while Fig.~\ref{fig:spnov} displays the Fourier transforms of the light curves of each of these stars along with the $0.1\%$\,FAP significance thresholds.

\textbf{MCT\,2148$-$2911} (TIC\,053851007, $G=16.07$\,mag,
$\alpha_{2000}=21^{\mathrm h}51^{\mathrm m}40^{\mathrm s}$,
$\delta_{2000}=-28^{\mathrm d}56^{\mathrm m}53^{\mathrm s}$),
also referred to as WD\,2148-291, was discovered to be a ZZ~Ceti variable by \citet{2006AJ....132..831G}. From a 1-hour light curve obtained in the $B$-band, they detected a single short-period signal of 260.8\,s with an amplitude of 12.6\,mma.  We identify no signals in the periodogram of the \textit{TESS} Sector 1 light curve that exceed the 0.1\% FAP significance threshold of 6.19\,mma, and specifically see no compelling peaks at the frequency measured by \citet{2006AJ....132..831G}.  This is not surprising, considering that the 2-minute \textit{TESS} exposures reduce the measured amplitude of a signal with this period to 69\% of its intrinsic amplitude, and the effect of observing in a redder bandpass reduces the amplitude again by a comparable factor (Section~\ref{sec:ross}).  Worse still, 83\% of the flux in the TESS aperture comes from other blended sources (CROWDSAP = 0.17), causing the significance threshold for this target to be 2.8 times higher than we would expect for an unblended source of the same magnitude.

\textbf{HE\,0031$-$5525} (TIC\,281594636, $G=15.80$\,mag,
$\alpha_{2000}=00^{\mathrm h}33^{\mathrm m}36^{\mathrm s}$,
$\delta_{2000}=-55^{\mathrm d}08^{\mathrm m}39^{\mathrm s}$) has a relatively low surface gravity for a ZZ~Ceti variable, with $\log{g} = 7.65\pm0.02$ determined by \citet{2006A&A...450..227C} from fitting hydrogen atmosphere models to the Balmer lines observed in its spectrum from the Hamburg ESO survey. However, \textit{Gaia} DR2 observations are consistent with a higher surface gravity, $\log{g} = 7.92\pm0.02$ \citep{2019MNRAS.482.4570G}. 
From five nights of time series photometry from the 1.6-m and 0.6-m telescopes at Observat{\'o}rio Pico dos
Dias, LNA, in Brazil, they detected three pulsation modes dominated by a 4.8\,mma signal with a period of 276.9\,s. This star was observed by \textit{TESS} in Sector 2 as TIC\,281594636, but with a significance threshold of 5.47\,mma (particularly high for a target of this magnitude due to blending of nearby sources), we do not detect any pulsation signals from this data set.

\textbf{EC\,00497$-$4723} (TIC\,101916028, $G=16.51$\,mag,
$\alpha_{2000}=00^{\mathrm h}52^{\mathrm m}01^{\mathrm s}$,
$\delta_{2000}=-47^{\mathrm d}07^{\mathrm m}09^{\mathrm s}$) was reported as a ZZ~Ceti pulsator in conference proceedings by \citet{1997fbs..conf..497S} based on multiple nights of both photoelectric and CCD observations.  A detailed analysis of these data is not presented, though the dominant pulsations in the 500--1000\,s range appear to vary in amplitude between roughly 10--20\,mma depending on the night. This star was observed by \textit{TESS} in Sector 2, with no significant signals detected above our 8.35\,mma threshold. The roughly day-timescale variations in pulsation amplitudes that were observed from the ground could significantly reduce their amplitudes measured in the \textit{TESS} periodogram compared to the average instantaneous pulsation amplitudes as power is distributed across multiple frequency bins.

\textbf{MCT\,0016$-$2553} (TIC\,246821917, $G=15.89$\,mag,
$\alpha_{2000}=00^{\mathrm h}18^{\mathrm m}45^{\mathrm s}$,
$\delta_{2000}=-25^{\mathrm d}36^{\mathrm m}43^{\mathrm s}$) was also discovered to pulsate by \citet{2006AJ....132..831G}. They detected a single pulsation signal with a 1152.4\,s period and an amplitude of 8.1\,mma in white light.  We do not detect any significant signals in the \textit{TESS} Sector 2 light curve of MCT\,0016$-$2553 to a threshold of 4.92\,mma, and we lose a lot of sensitivity because of blending of nearby sources within the TESS aperture (CROWDSAP = 0.13).

\textbf{WD\,0108$-$001} (TIC\,336891566, $G=17.82$\,mag,
$\alpha_{2000}=01^{\mathrm h}11^{\mathrm m}24^{\mathrm s}$,
$\delta_{2000}=+00^{\mathrm d}09^{\mathrm m}35^{\mathrm s}$) was reported as SDSS\,J0111+0009 to be a ZZ~Ceti variable in a wide binary system with an M dwarf companion by \citet{2015MNRAS.447..691P}. This target happened to fall within the field of \textit{K2} Campaign 8 and was observed for 78.72 days with a 1-minute cadence. \citet{2017ApJS..232...23H} detected 14 frequencies associated with seven independent pulsation modes from this data set, though the largest amplitude measured from the \textit{K2} light curve reached only 3.88\,mma---well below the noise level of the \textit{TESS} Sector 3 periodogram (0.1\% FAP threshold is 10.93\,mma).

\textbf{HS\,0235+0655} (TIC\,365247111, $G=16.51$\,mag, 
$\alpha_{2000}=02^{\mathrm h}38^{\mathrm m}33^{\mathrm s}$,
$\delta_{2000}=+07^{\mathrm d}08^{\mathrm m}09^{\mathrm s}$) was discovered to pulsate by \citet{2007ASPC..372..583V}. They report a single main periodicity of 1283.7\,s with an amplitude of 4.21\,mma, though the 1.2-hr light curve segment they display exhibits the characteristic amplitude modulation caused by beating of closely spaced modes.  The periodogram of the \textit{TESS} Sector 4 light curve achieved a very low signal-to-noise ratio, with a 0.1\% FAP threshold of 12.42\,mma.

\textbf{KUV\,03442+0719} (TIC\,468887063, $G=16.62$\,mag,
$\alpha_{2000}=03^{\mathrm h}46^{\mathrm m}51^{\mathrm s}$,
$\delta_{2000}=+07^{\mathrm d}28^{\mathrm m}03^{\mathrm s}$) was discovered to pulsate with an amplitude of 7.6\,mma at a dominant period of 1384.9\,s by \citet{2006AJ....132..831G}. Curiously, \citet{2007ASPC..372..583V} did not detect any pulsations from this star to a limit of 2\,mma.  Unsurprisingly, no signals exceed the 16.62\,mma detection threshold of the \textit{TESS} Sector 5 data.

\textbf{WD\,J0925+0509} (TIC\,290653324, $G=15.27$\,mag, 
$\alpha_{2000}=09^{\mathrm h}25^{\mathrm m}12^{\mathrm s}$,
$\delta_{2000}=+05^{\mathrm d}09^{\mathrm m}33^{\mathrm s}$) found to be a massive white dwarf of $0.87\pm0.01M_{\odot}$ that was discovered to pulsate by \citet{2010MNRAS.405.2561C}, whom detected two signals with periods 1127.14 and 1264.29\,s with low amplitudes of 3.17 and 3.05\,mma, respectively.
However, \textit{Gaia} DR2 observations suggest a moderate mass of $M_* = 0.74 \pm 0.01\,M_\odot$ \citep{2019MNRAS.482.4570G}.
The \textit{TESS} Sector 8 data are only sensitive to signals that exceed a 3.81\,mma threshold, of which we find none.

\textbf{HS\,1013+0321} (TIC\,277747736, $G=15.69$\,mag, 
$\alpha_{2000}=10^{\mathrm h}15^{\mathrm m}48^{\mathrm s}$,
$\delta_{2000}=+03^{\mathrm d}06^{\mathrm m}48^{\mathrm s}$) is a ZZ~Ceti near the hot edge of the instability strip that was detected to pulsate with three modes of periods (amplitudes) 270.0\,s (8.4\,mma), 255.7\,s (7.3\,mma), and 194.7\,s (5.8\,mma) by \citet{2004ApJ...607..982M}. Continued monitoring of this star by \citet{2008ApJ...676..573M} refined the measurement of the second period to $254.918450\pm0.000006$\,s, and constrained the rate of change of the period to $\dot{P} = (7.2\pm3.6) \times 10^{13}$\,s\,s$^{-1}$.  The observable amplitudes of these short period modes would all be significantly decreased by \textit{TESS}'s 2-minute exposures, and none appear to exceed the significance threshold of 4.59\,mma from the Sector 8 light curve.

\textbf{EC\,11266$-$2217} (TIC\,219442838, $G=16.42$\,mag,
$\alpha_{2000}=11^{\mathrm h}29^{\mathrm m}12^{\mathrm s}$,
$\delta_{2000}=-22^{\mathrm d}33^{\mathrm m}44^{\mathrm s}$) was discovered to exhibit at least four pulsation signals between 215--403\,s by \citet{2006A&A...450.1061V}, with the highest peak reaching an amplitude of 7\,mma. The \textit{TESS} Sector 9 data do not reveal any signals to exceed a 5.20\,mma detection limit. The light curve does exhibit two moments of significant brightness enhancement during days 2\,458\,545 and 2\,458\,548 BJD, the first reaching as high as 2.8 times the mean flux level in the processed light curve (Figure~\ref{fig:lcnov}). These are caused by stellar flares of the nearby M-dwarf star that contributes most of the flux to the TESS photometric aperture \citep[the mean contaminating flux level of 65\% was subtracted from the light curve by the PDC pipeline;][]{2010SPIE.7740E..1UT}. Astrometry from \textit{Gaia} DR2 indicates that this M dwarf (\verb+source_id+ 3541237717085786880) is a common-proper-motion companion to the ZZ~Ceti variable (\verb+source_id+ 3541237717085787008).

\end{appendix}

\end{document}